\documentclass[letterpaper,twocolumn,10pt]{article}
\usepackage{usenix2019_v3}
\usepackage{cite}
\usepackage{amsmath,amssymb,amsfonts}
\usepackage{algorithmic}
\usepackage{graphicx}
\usepackage{textcomp}
\usepackage{xcolor}
\usepackage{pifont}
\usepackage{subfigure}
\usepackage[symbol]{footmisc}

\usepackage{tikz}
\usepackage{amsmath}
\usepackage{url}
\usepackage[compact]{titlesec}

\usepackage{filecontents}

\newcommand{\para}[1]{\noindent {\bf #1}}

\titlespacing{\section}{0pt}{1ex}{0ex}
\titlespacing{\subsection}{0pt}{0.5ex}{0ex}
\titlespacing{\subsubsection}{0pt}{0.5ex}{0ex}
\begin{document}

\date{}

\title{\Large \bf A Comprehensive Study on Optimizing Systems with Data Processing Units}

\author{
{\rm Shangyi Sun, Chunpu Huang, Rui Zhang, Lulu Chen, Yukai Huang, Ming Yan, Jie Wu}\\
Fudan University
} 

\maketitle

\begin{abstract}

New hardware, such as SmartNICs, has been released to offload network applications in data centers. Off-path SmartNICs, a type of multi-core SoC SmartNICs, have attracted the attention of many researchers. Unfortunatelly, they lack the fully exploration of off-path SmartNICs. 

In this paper, we use a BlueField SmartNIC as an example to conduct a systematical study on the advantages and disadvantages of off-path SmartNICs. We make a detailed performance characterization on an off-path SmartNIC including computing power and network communication overhead, and propose the following advices: 1) Directly utilize the specific accelerators on the SmartNIC to offload applications; 2) Offload latency-insensitive background processing to the SmartNIC to reduce the load on the host; 3) Regard the SmartNIC as a new endpoint in the network to expand the computing power and storage resources of the server host; 4) Avoid directly employing the design method for systems based on on-path SmartNICs. We apply these advices to several use cases and show the performance improvements.
\end{abstract}

\section{Introduction}
When storage and computing resources cannot meet current performance requirements, traditional data centers often need to add new servers. Since the physical space of the data center is limited, this method sometimes cannot meet the actual demand. Therefore, researchers are constantly exploring other ways to improve system performance. For example, many studies have found that data transmission in distributed systems often brings a lot of overhead and even becomes the bottleneck of the system, so Remote Direct Memory Access (RDMA) is used to improve the performance of distributed applications\cite{huang2021nova,farm,pilaf,wei2015fast,chen2016fast,lu2017octopus,yang2019orion,aguilera2018remote,behrens2018rdmc,kalia2018datacenter,trivedi2015rstore,stuedi2014darpc,tsai2017lite}. However, using RDMA to accelerate systems can only offload simple operations such as reading and writing memory to the NIC\cite{schuh2021xenic}. It cannot offload more complex tasks. Now, SmartNICs give us new ways to perform more complex offloading, which can expand the computing and storage resources of servers at a low cost.

Over the past few years, many hardware manufacturers have released different types of multi-core SoC (system-on-a-chip) SmartNICs, including Mellanox BlueField\cite{BF}, Marvell LiquidIO\cite{LiquidIO}, Broadcom Stingray\cite{Stingray}, and Agilio Netronome\cite{Agilio}. They can be divided into two categories according to whether the core of the SmartNIC is on the data transmission path: on-path multi-core SoC SmartNICs (on-path SmartNICs) and off-path multi-core SoC SmartNICs (off-path SmartNICs)\cite{liu2019offloading}. Earlier studies are mostly based on on-path SmartNICs, which offload tasks such as network functions and packet scheduling\cite{Floem,liu2019offloading,liu2019e3,qiu2021automated,schuh2021xenic,shashidhara2022flextoe,gao2021ovs}. In recent years, there have been more and more researches on off-path SmartNICs. Different from on-path SmartNICs, the NIC switch on off-path SmartNICs can choose to forward the received data packet directly to the host or the core on the NIC for further processing. In addition, they tend to have a complete network stack and full operating system, providing high programmability. Therefore, in some recent studies, off-path SmartNICs are used to offload file system\cite{kim2021linefs}, network address translation\cite{fang2021hypernat}, key-value store\cite{sun2022skv}, etc.

Although off-path SmartNICs have many advantages over on-path SmartNICs, using them to offload applications can be challenging. These challenges come from many aspects. First, the applications are complex so developers face many choices when offloading. Second, the resources in SmartNIC are limited. For example, the processor cores on SmartNICs are weaker than host CPU cores. Third, applications always have strict requirements on performance, scalability, availability, etc. 
As a result, developers have to take many factors into consideration when offloading.

To make full and effective use of a SmartNIC and choose the most appropriate offloading method, it is necessary to have a thorough understanding of its structure and performance. Therefore, this paper first analyzes the internal structure and different modes of operation of a BlueField SmartNIC. Then we use some micro-benchmarks to evaluate the computing power and communication overhead of the SmartNIC, which has not been thoroughly studied in existing works. In addition to verifying the conclusions of some previous studies, we obtain some new findings in our experiments.

Our evaluation results show that despite the high programmability of the off-path SmartNICs, they still have some disadvantages which pose significant challenges to the development of SmartNIC-based applications. We discover that the same task is performed much slower on the SmartNIC than on the host, and the communication latency between the SmartNIC and the host is very high. Due to these drawbacks, offloading applications arbitrarily with off-path SmartNICs can result in a decrease in overall performance. Only by making full use of the advantages of the SmartNIC and avoiding its shortcomings can it play its role to the greatest extent. Based on the characteristics of off-path SmartNICs, we propose the following design advices:
\begin{itemize}
\setlength{\itemsep}{0pt}
\setlength{\parsep}{0pt}
\setlength{\parskip}{0pt}
\item Accelerate applications with the specific accelerators on the SmartNIC. Off-path SmartNICs are often equipped with special accelerators, such as encryption and decryption accelerators and regular expression matching accelerator. These accelerators use special hardware to offload processing logic and can achieve better performance compared with host-based methods. When developing an application, the accelerators can be called directly through user interfaces.
\item Offload latency-insensitive background operations. Many systems need to perform some background processing in addition to constantly interacting with clients. For example, in a file system, log processing is a background operation. Such an operation has no explicit latency requirements, so offloading it to weaker SmartNIC cores will not affect the overall performance. Offloading such operations reduces the load on the host, allowing the host to respond to clients faster.
\item Treat the SmartNIC as a new endpoint to expand the computing power and storage space of the server host. This is because the off-path SmartNICs have multi-core processors, DRAM and non-volatile storage media, which is similar to general hosts. In addition, an off-path SmartNIC usually has a complete network stack such as RDMA and an independent IP address, so it can be used as an independent node in the network.
\item Avoid directly employing the design method of systems based on on-path SmartNICs. Most of the current researches for multi-core SoC SmartNICs are based on on-path SmartNICs, but it is not feasible to directly adopt their methods when using an off-path SmartNIC. It is because, in some aspects, off-path SmartNICs perform much worse than on-path SmartNICs. Ignoring these drawbacks will make the overall performance of the system degrade significantly.
\end{itemize}

We also explain the rationale of these advices in detail. After that, we conduct several experiments to demonstrate the feasibility of these advices. Experimental results show that the performance of the system is improved obviously by applying these advices.

The next several sections are organized as follows. Section \ref{sec2} introduces the background of the off-path SmartNICs, especially the difference between off-path SmartNICs and on-path SmartNICs, and analyzes the structure of a Bluefield SmartNIC. Section \ref{sec3} characterizes the performance of the SmartNIC and points out design advices based on SmartNIC structure and performance characterization. Section \ref{seccase} illustrates the feasibility of the design advices through some use cases. Section \ref{secrelated} summarizes the related works and discusses some of them according to our advices. Section~\ref{sec:conclusion} summarizes our contribution and the design advices proposed in this paper.

\section{Background and Motivation}
\label{sec2}
This section introduces the background of SmartNICs and highlights the differences between off-path SmartNICs and on-path SmartNICs. Then, we take Mellanox BlueField as an example to describe the internal structure of an off-path SmartNIC, which is the basis of our design guidelines.

\subsection{Differences between off-path and on-path SmartNICs}
Unlike FPGA-based SmartNICs\cite{Tonic,li2016clicknp,ibanez2021nanopu,lin2020panic,eran2019nica,firestone2018azure,li2017kv,lavasani2013fpga,lim2013thin,tokusashi2016multilevel}, multi-core SoC SmartNICs do not need to consolidate application logic into the programmable logic blocks of the NIC, which makes it easier to offload complex processing logic. These SmartNICs have multi-core processors in addition to on-board DRAM. Based on whether the cores of the SmartNIC are on the packet transmission path, multi-core SoC SmartNICs can be divided into two types: on-path and off-path SmartNIC. Each of them has its advantages and disadvantages.

\begin{table}[htbp] 
\caption{Characteristics of on-path and off-path SmartNICs}
\label{tab:threesome}
\centering
\setlength{\tabcolsep}{4mm}
\begin{tabular}{lll}
\hline
     & On-path & Off-path \\
\hline
    NIC switch & \ding{56} & \ding{52} \\
    OS & \ding{56} & \ding{52} \\
    Network stack  &  \ding{56} & \ding{52} \\
    Low-level interfaces & \ding{52} & \ding{56} \\
    Communication overhead & low & high \\
    
\hline   
\end{tabular}
\end{table}

 The characteristics of on-path and off-path SmartNICs are shown in Table \ref{tab:threesome}.In addition to the components such as the on-board memory and processors, an off-path SmartNIC is also equipped with a NIC switch. The NIC switch decides whether to forward the incoming packets to the NIC cores or to the host where the SmartNIC is located. This does not require a packet to pass through the core of the NIC compulsively, which is different from on-path SmartNICs. In addition, unlike on-path SmartNICs which only provide low-level interfaces for manipulating packets, off-path SmartNICs typically run with a full network stack and operating system, making them more programmable than on-path SmartNICs. Although on-path SmartNICs have powerful data flow processing capabilities, they have small storage space, and lack timers and support for floating-point and other complex computations such as division\cite{shashidhara2022flextoe}. In contrast, off-path SmartNICs support more complex offloading.

However, off-path SmartNICs also have shortcomings. They do not provide low-level interfaces for manipulating packets, which makes it impossible for developers to directly process incoming and outgoing packets at will. In addition, because an off-path SmartNIC has a complete network protocol stack, the communication cost between the SmartNIC cores and the host is very high compared with on-path SmartNICs. Unlike an on-path SmartNIC, which provides low-level interfaces for host memory manipulation, off-path SmartNICs can only communicate with the host through protocols such as RDMA or TCP. The experimental results in Section \ref{secperf} show that a BlueField2 has a huge communication overhead and previous works\cite{schuh2021xenic,xing2022towards} show that other off-path SmartNICs such as Broadcom Stingray face the same problem. A simple offload can result in significant performance degradation compared with host-based implementation. These shortcomings make it challenging to develop applications based on off-path SmartNICs.

\subsection{BlueField SmartNIC}
BlueField is a typical off-path SmartNIC. A BlueField2 SmartNIC contains 8 ARM AArch64 processors, which are highly programmable. It is equipped with its own software ecosystem that aims to provide high-speed networking, storage, and security services.

\begin{figure}[htbp]
\centerline{\includegraphics[width=0.4\textwidth,trim=140 50 100 0,clip]{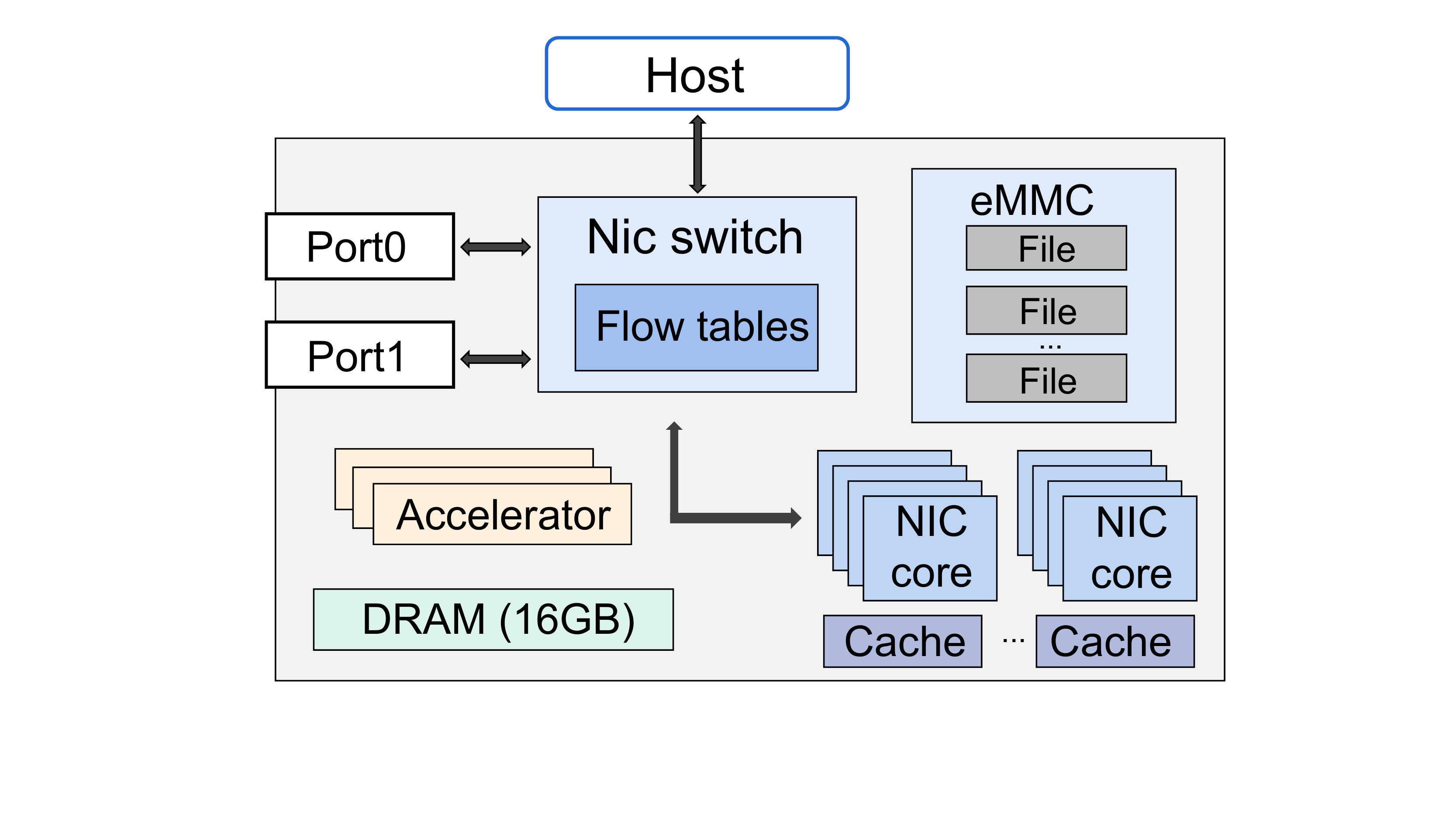}}
\caption{The architecture of a BlueField SmartNIC.}
\label{bf2arc}
\end{figure}


A BlueField SmartNIC has three types of network interfaces: ConnectX Ethernet/InfiniBand interface, RShim (Random Shim) virtual Ethernet interface and OOB (Out-of-band) Ethernet interface. The ConnectX Ethernet/InfiniBand interface provides high-speed network communication, which is suitable for production environments. The RShim virtual Ethernet interface can be used for debugging, installation or basic management. A special driver called RShim must be installed and run to expose the various BlueField management interfaces on the host operating system. This virtual network interface can only run at about 10MB/s and should not be used for network transmission in production environments. A BlueField SmartNIC also has an OOB 1Gb management port. This interface supports TCP/IP network connections to Arm cores but it cannot boot the SmartNIC like the RShim interface.

The internal hardware structure of BlueField MBF2H516A SmartNIC is shown in Figure \ref{bf2arc}. A BlueField SmartNIC has 8 ARM cores and every two cores share a 1MB L2cache. It has 16GB on-board DRAM. Although it is smaller than the DRAM on the host, it is enough to do some more complex processing than on-path SmartNICs. It also has eMMC flash memory, which provides storage that will not be lost after a power failure. It has a NIC switch. Based on the flow table on it, the NIC switch can decide whether to forward incoming packets to the NIC core or the host core. A very important feature of BlueField is that it is equipped with many dedicated accelerators, including encryption/decryption and regular expression matching accelerators. Better performance can be achieved by offloading applications to these hardware accelerators. BlueField provides the DOCA framework, and users can call its interfaces to access the accelerators conveniently.

BlueField provides three operation modes: embedded function mode, restricted mode and separated host mode. Switching between modes is implemented by configuring processing rules on the NIC switch. Embedded function mode is the default mode. In this mode, all resources in BlueField are managed by the ARM cores. All network traffic to and from the host goes through the embedded ARM cores. This mode is suitable for offloading a virtual network switch to the SmartNIC, which can filter, direct, monitor and apply QoS rules to packets traveling between the host and the network. BlueField in this mode is a bit like an on-path SmartNIC, but it lacks low-level interfaces to manipulate data packets. Restricted mode is a special embedded functional mode that sets additional restrictions on host access to the SmartNIC for security purposes. Separated host mode is the most commonly used mode and many existing works are based on this mode. Under this mode, the SmartNIC acts just like a new endpoint in the network. Due to the complete network stack, the SmartNIC can communicate with the local host and the remote host through Ethernet or RoCE (RDMA over Converged Ethernet)\cite{ROCE}.

\subsection{Motivation}
Among existing works, some of them have started offloading applications to off-path SmartNICs without understanding the features of the SmartNICs\cite{min2021gimbal,kim2021linefs,tork2020lynx,fang2021hypernat}. There are also researchers that characterize the performance of off-path SmartNICs\cite{liu2021performance}. However, their performance tests for SmartNICs are not comprehensive, and they do not summarize any guidelines for using off-path SmartNICs. In addition, most of the works tend to focus on offloading a specific application, such as file system\cite{kim2021linefs}, distributed key-value store\cite{sun2022skv}, and network address translation\cite{fang2021hypernat}. However, none of them has proposed a universal solution for using off-path SmartNICs for any systems. Therefore, in this paper, we hope to summarize a series of design guidelines based on the performance characterization and internal structure of the SmartNIC to make it easier for developers to build systems using off-path SmartNICs.

\section{Performance Characterization and Design Guidelines}
\label{sec3}
To better leverage the strength of off-path SmartNIC, we first make a detailed performance characterization of a BlueField SmartNIC. Then we propose 4 design guidelines on how to build systems when using an off-path SmartNIC.

\subsection{Performance Characterization}
\label{secperf}
Here, we first evaluate the computing power of the SmartNIC compared with the host, and then we test the overhead of memory access and network communication.

\subsubsection{Experimental Setup}
Each server in our experiment has two 16-core Intel(R) Xeon(R) Gold 5218 CPU processors with the frequency of 2.30 GHz, and the size of DRAM is 64 GB. Each CPU core is equipped with a 64 KB L1 cache and a 1 MB L2 cache. 16 cores on a CPU share a 22 MB L3 cache. Our servers have Ubuntu 18.04 operating systems on them. The configuration of the BlueField MBF2H516A SmartNIC used in our experiments is shown in Section \ref{sec2}.

\subsubsection{Computing Power}
In order to know the computing power of a SmartNIC compared with a host, we evaluate it through various micro-benchmarks. We choose to use \texttt{stress-ng}\cite{Stressng} to evaluate the performance of the SmartNIC and the host. The \texttt{stress-ng} tool has a large number of testing functions called stressors. These stressors cover a variety of executable tasks and they can reflect the performance of all aspects of the processor thoroughly and meticulously, including IO, CPU operations, file systems and system calls. All stressors are divided into several classes. There are multiple stressors in each class, and each stressor belongs to one or more classes. The \texttt{stress-ng} tool can perform various tests flexibly, such as repeatedly executing a system call within a specified period of time. In addition, \texttt{stress-ng} can specify the number of workers executing the task, which demonstrates the scalability. Running the stressors on the SmartNIC can show the computing power of the general-purpose ARM cores on it.
\begin{figure}
\begin{minipage}[t]{0.54\linewidth}
\includegraphics[width=\linewidth,trim=0 22 0 0,clip]{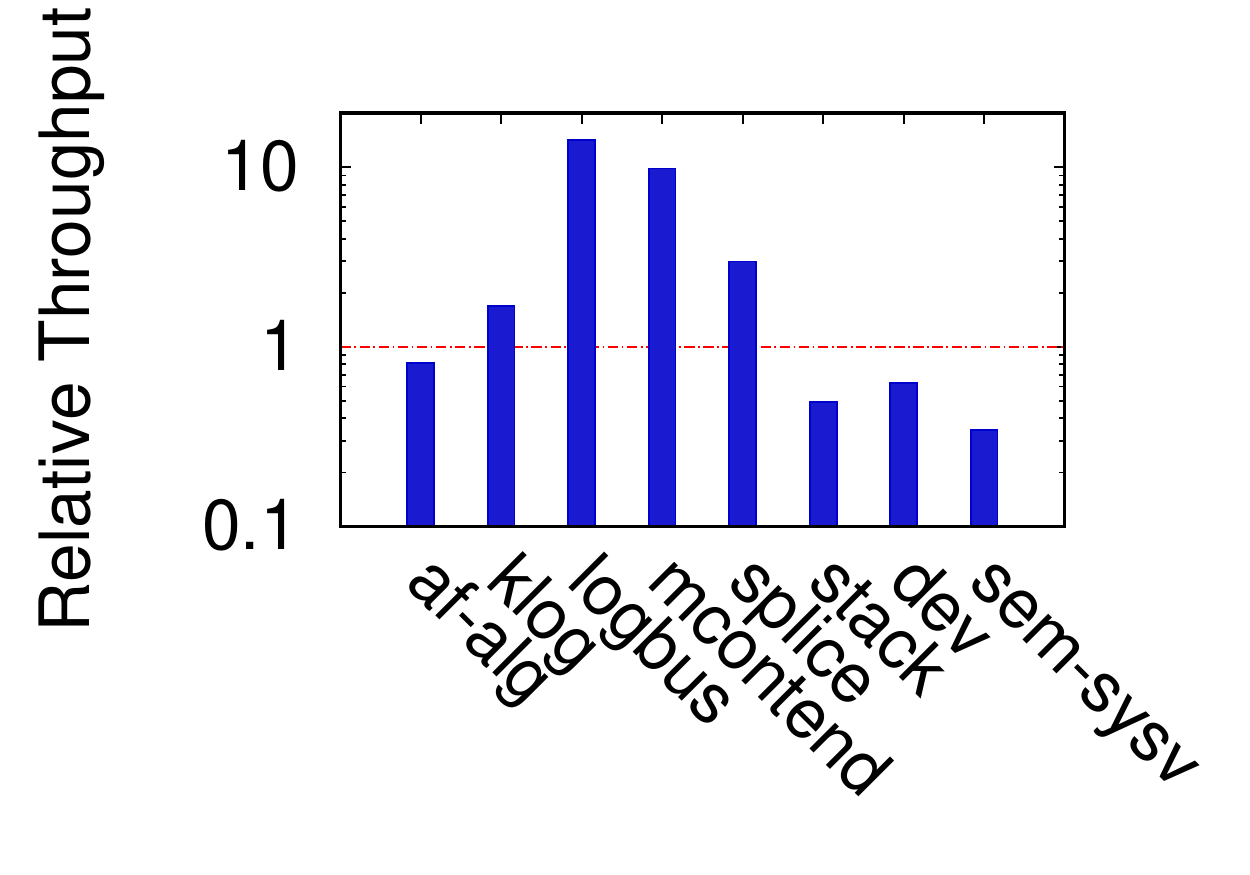}
\caption{The relative throughput of executing the stressors on SmartNIC.}
\label{stress1}
\end{minipage}%
\hfill%
\begin{minipage}[t]{0.43\linewidth}
\includegraphics[width=\linewidth,trim=30 0 30 0,clip]{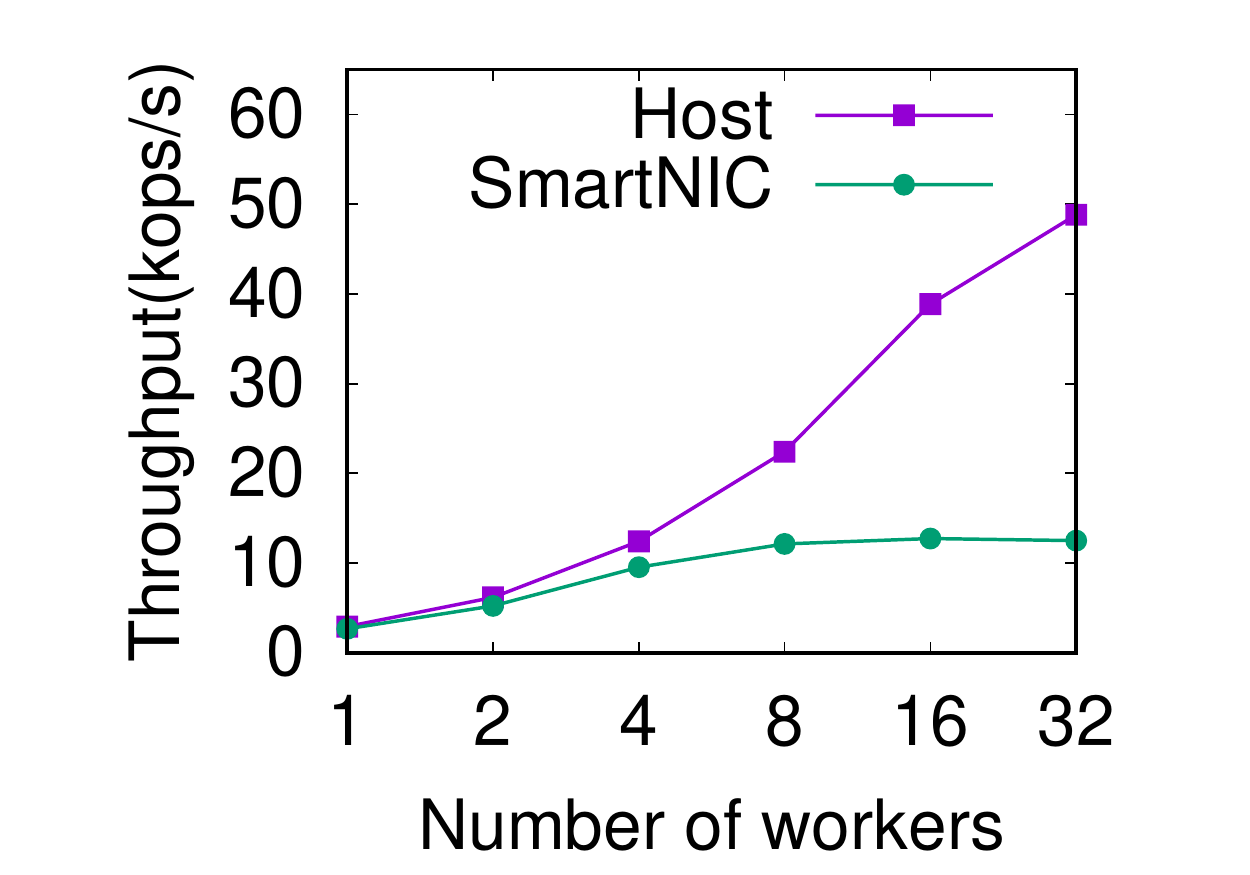}
\caption{The scalability of af-alg stressor on host and SmartNIC.}
\label{stress2}
\end{minipage}
\end{figure}

Previous work\cite{liu2021performance} compares the results of a BlueField SmartNIC running \texttt{stress-ng} against 13 ageing servers, and shows the ranking of BlueField in all 14 cases when executing each stressor. BlueField ranks first (af-alg,klog,lockbus,mcontend,splice,stack) or second (dev,sem-sysv) among these 14 test platforms in eight stressor tests. We also execute these stressors on our host and SmartNIC. The results are shown in Figure \ref{stress1}. The relative throughput equals the throughput of SmartNIC divided by the throughput of host. Only 4 stressors (klog, lockbus, mcontent, splice) get better performance when executed on the SmartNIC compared with the host. On our experiment platform, the SmartNIC performs worse than the host when executing af-alg and stack, which is different from previous work. We infer this is due to the better performance of our host cores compared with the host in previous work.

\begin{table}[htbp] 
\caption{Throughput (bogo-ops-per-second) of running CPU-intensive stressors on the host and on the SmartNIC.}
\label{stresscpu}
\centering
\setlength{\tabcolsep}{6mm}
\begin{tabular}{lll}
\hline
    Stressor & Host & SmartNIC \\
\hline
    atomic & 181716.9 & 171942.31 \\
    branch & 124392.88 & 111940.98 \\
    bsearch & 385.46 & 303.64 \\
    context & 6360.07 & 2048.77 \\
    cpu & 1389.20 & 151.27 \\
    crypt & 1196.93 & 823.5 \\
    hash & 82835.08 & 35500.64\\
    heapsort & 3.87 & 2.5 \\
    goto & 250457.10 & 203355.43 \\
    matrix & 3396.54 & 1154.74 \\
    mergesort & 26.25 & 13.25 \\
    qsort & 12.13 & 3.37 \\
    skiplist & 6129.61 & 3726.68 \\
    str & 53560.45 & 22211.69 \\
    tree & 1.87 & 0.5 \\
    
\hline   
\end{tabular}
\end{table}

\begin{figure*}[!t]
\centering
\subfigure[Sequential read.]{
\begin{minipage}[t]{0.25\linewidth}
\centering
\includegraphics[width=1.9in,trim=0 0 0 0,clip]{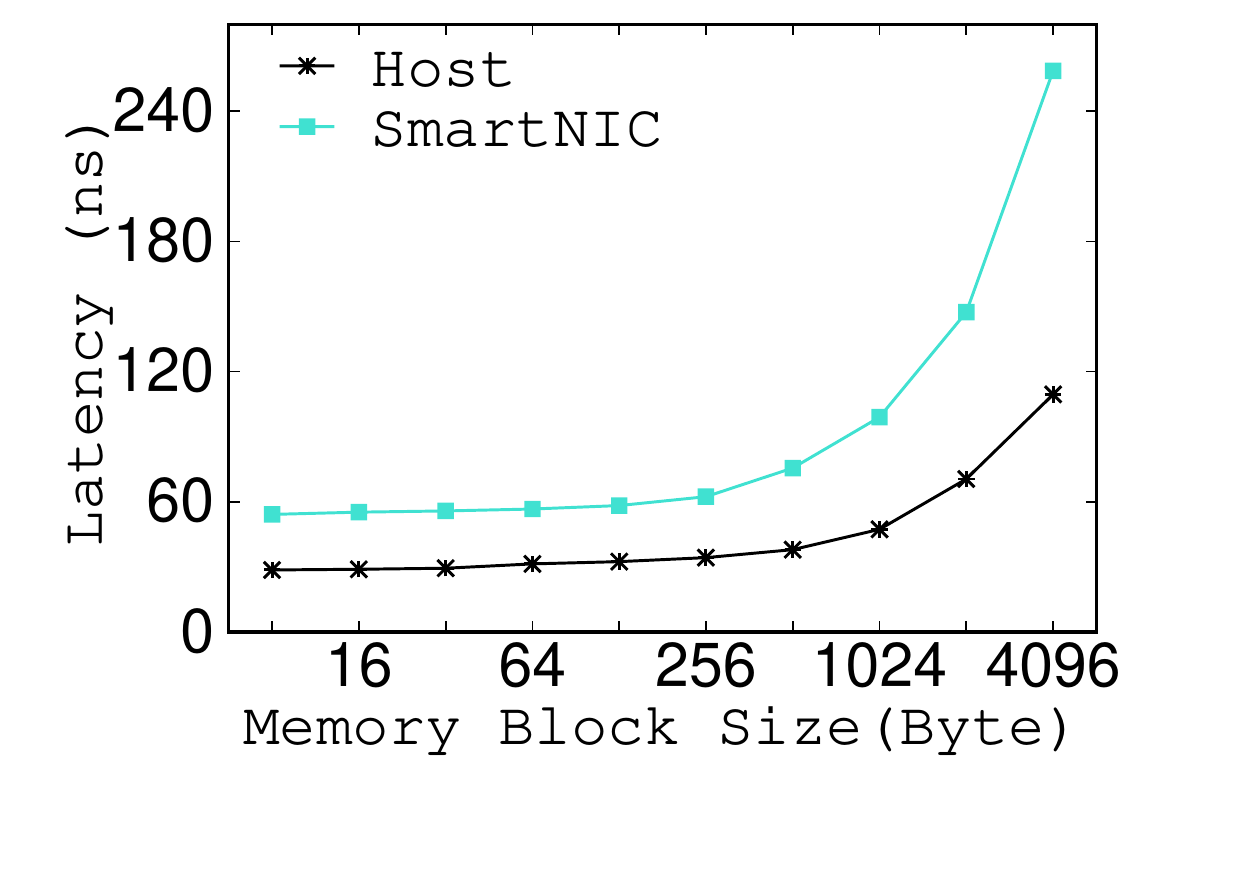}
\end{minipage}%
}%
\subfigure[Random read.]{
\begin{minipage}[t]{0.25\linewidth}
\centering
\includegraphics[width=1.9in,trim=0 0 0 0,clip]{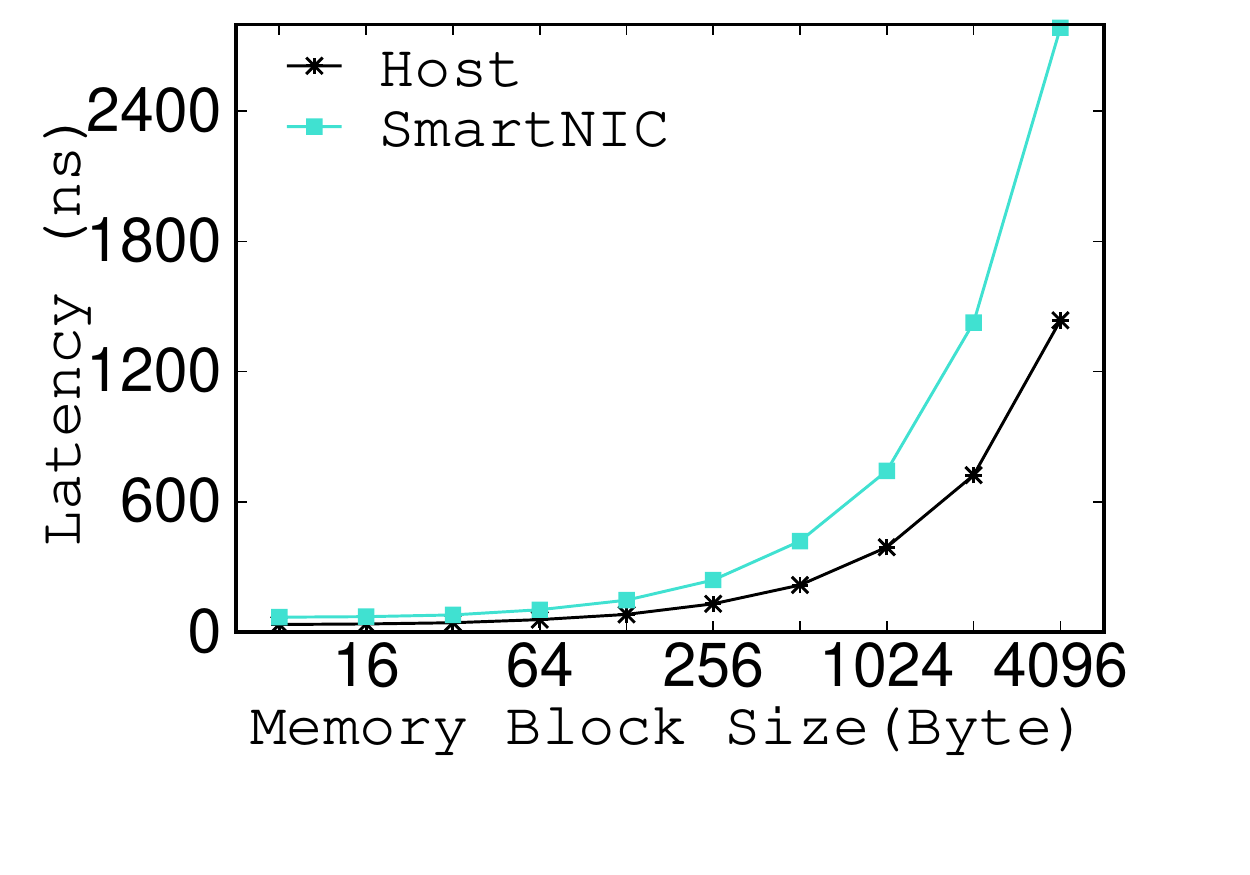}
\end{minipage}%
}%
\subfigure[Sequential write.]{
\begin{minipage}[t]{0.25\linewidth}
\centering
\includegraphics[width=1.9in,trim=0 0 0 0,clip]{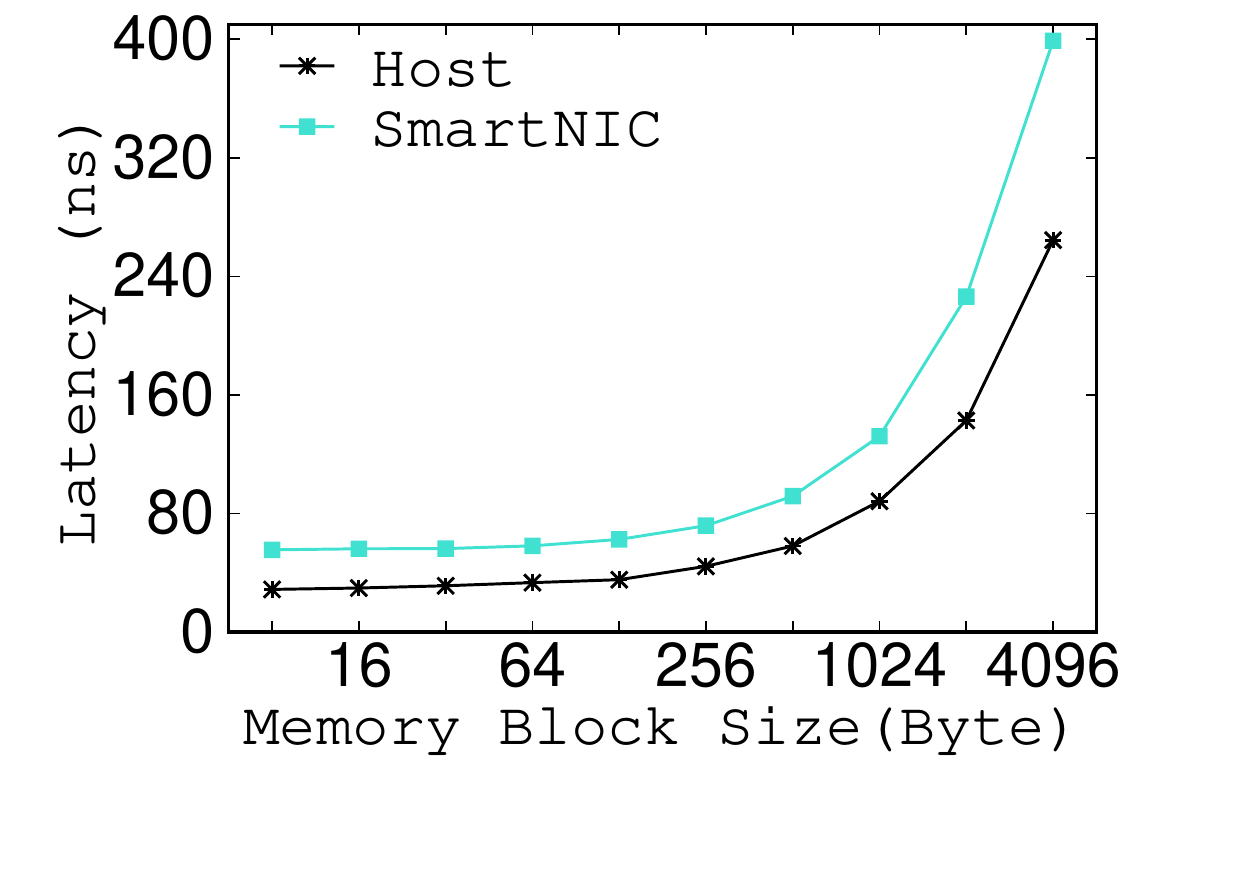}
\end{minipage}
}%
\subfigure[Random write.]{
\begin{minipage}[t]{0.25\linewidth}
\centering
\includegraphics[width=1.9in,trim=0 0 0 0,clip]{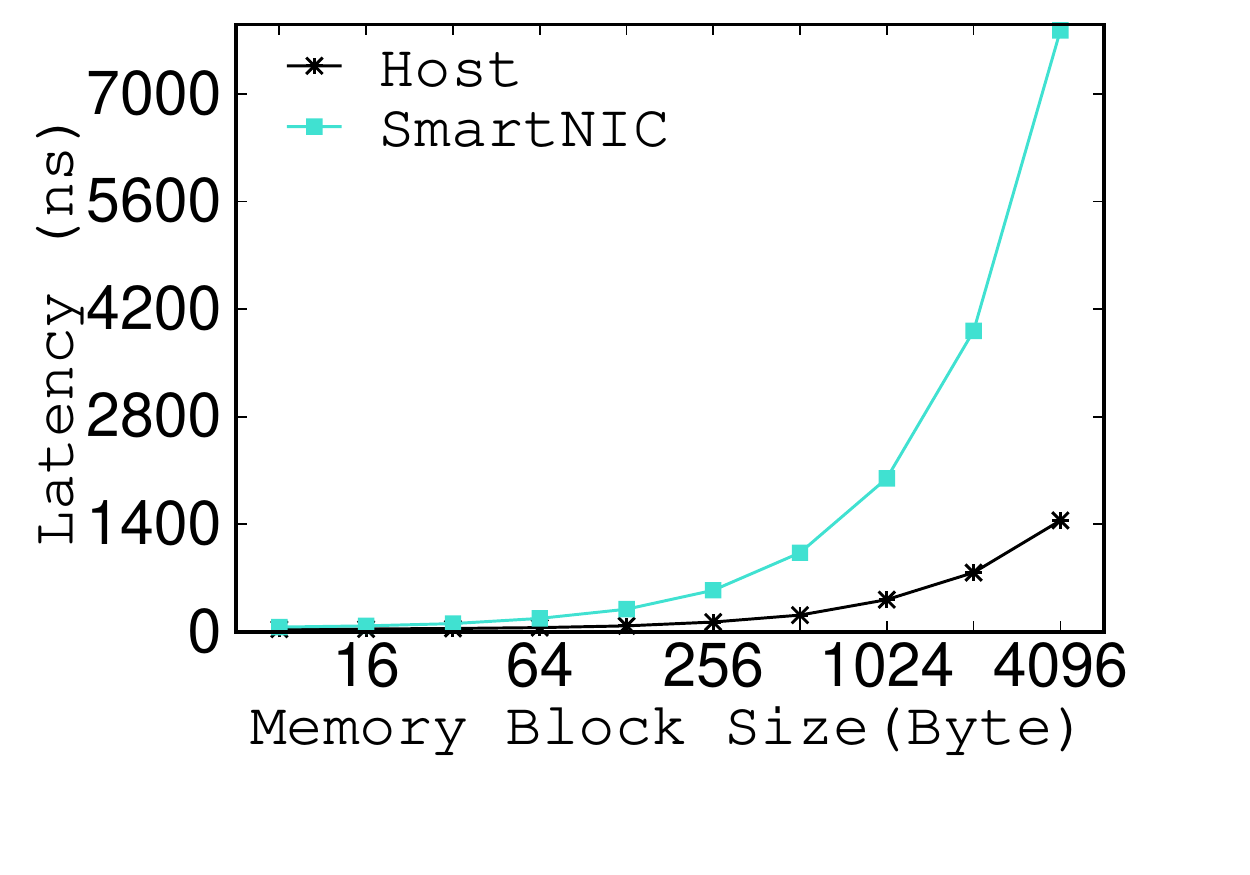}
\end{minipage}%
}%

\centering
\caption{The overhead of memory access of the SmartNIC and the host.}
\label{figmem}
\end{figure*}
To explore the performance of SmartNICs when handling CPU-intensive tasks, we executed some stressors of CPU class on the SmartNIC and the host. The results are shown in Table~\ref{stresscpu}. These stressors represent many common tasks executed on CPU. For example, mergesort means merge sorting 32-bit random integers, and bsearch means exercising a binary search. Note that the results are reported in terms of bogus operations per second (bogo-ops-per-second). Although this provides a way to compare the performance of executing the same task on different platforms, the value of bogo-ops-per-second of different stressors varies a lot. All stressors perform better when running on the host compared with running on the SmartNIC. This shows that the general-purpose ARM core of the SmartNIC is much weaker than the host core, which is consistent with the results reported in previous work\cite{liu2021performance}.

In addition, we test the scalability of a stressor running on the host and the SmartNIC. We take af-alg as an example and increase the number of workers from 1 to 32. The result is shown in Figure \ref{stress2}. In the case of 1 worker, the throughput of the host is only a little higher than the SmartNIC. However, as the number of workers increases, the gap between them becomes larger. The stressor shows worse scalability when running on SmartNIC, because the SmartNIC has fewer cores than the host. Moreover, a large number of worker threads preempt limited computing resources, resulting in the overhead of context switching. This also leads to worse performance of the SmartNIC in this case.

\begin{figure*}[!t]
\centering
\subfigure[RDMA Write latency.]{
\begin{minipage}[t]{0.32\linewidth}
\centering
\includegraphics[width=2.4in,trim=0 30 0 30,clip]{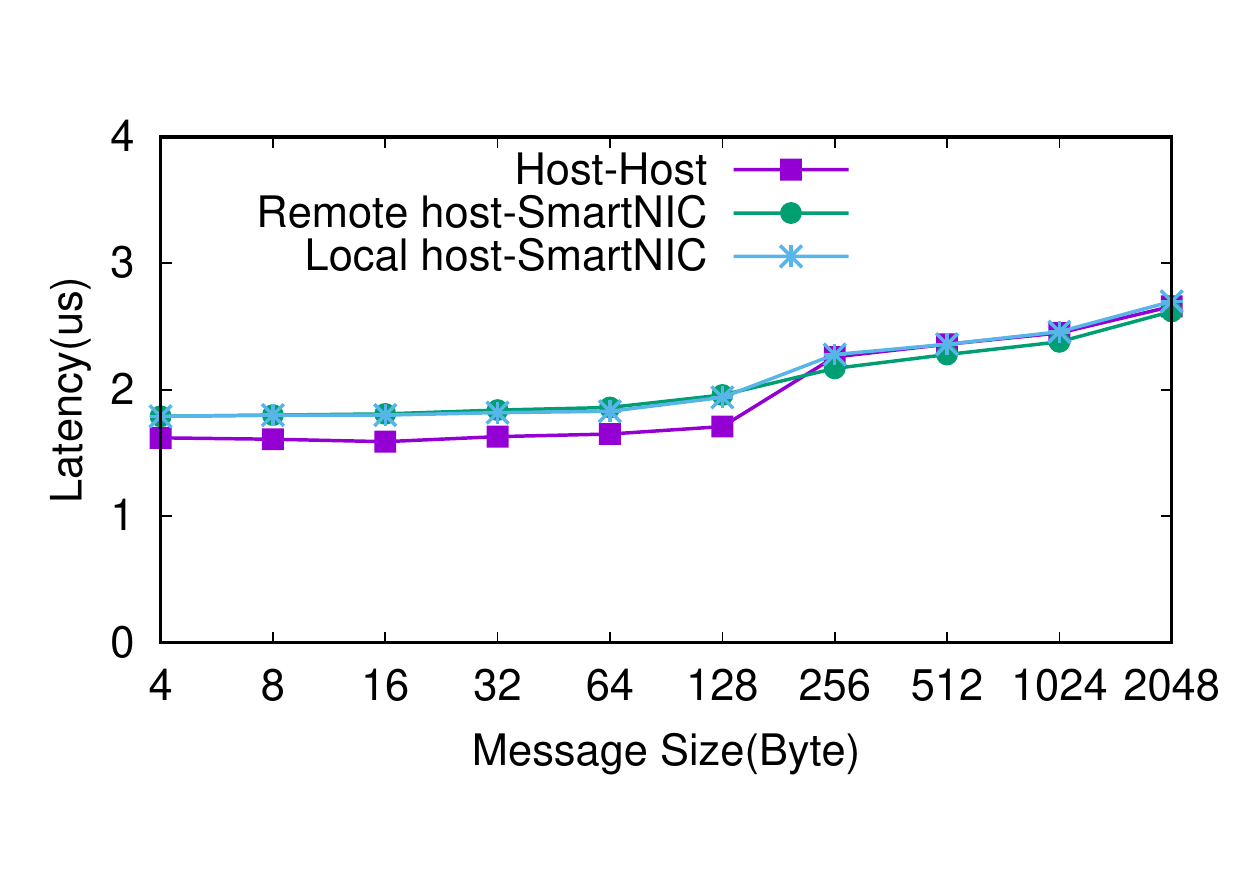}
\end{minipage}%
}%
\subfigure[RDMA Read latency.]{
\begin{minipage}[t]{0.32\linewidth}
\centering
\includegraphics[width=2.4in,trim=0 30 0 30,clip]{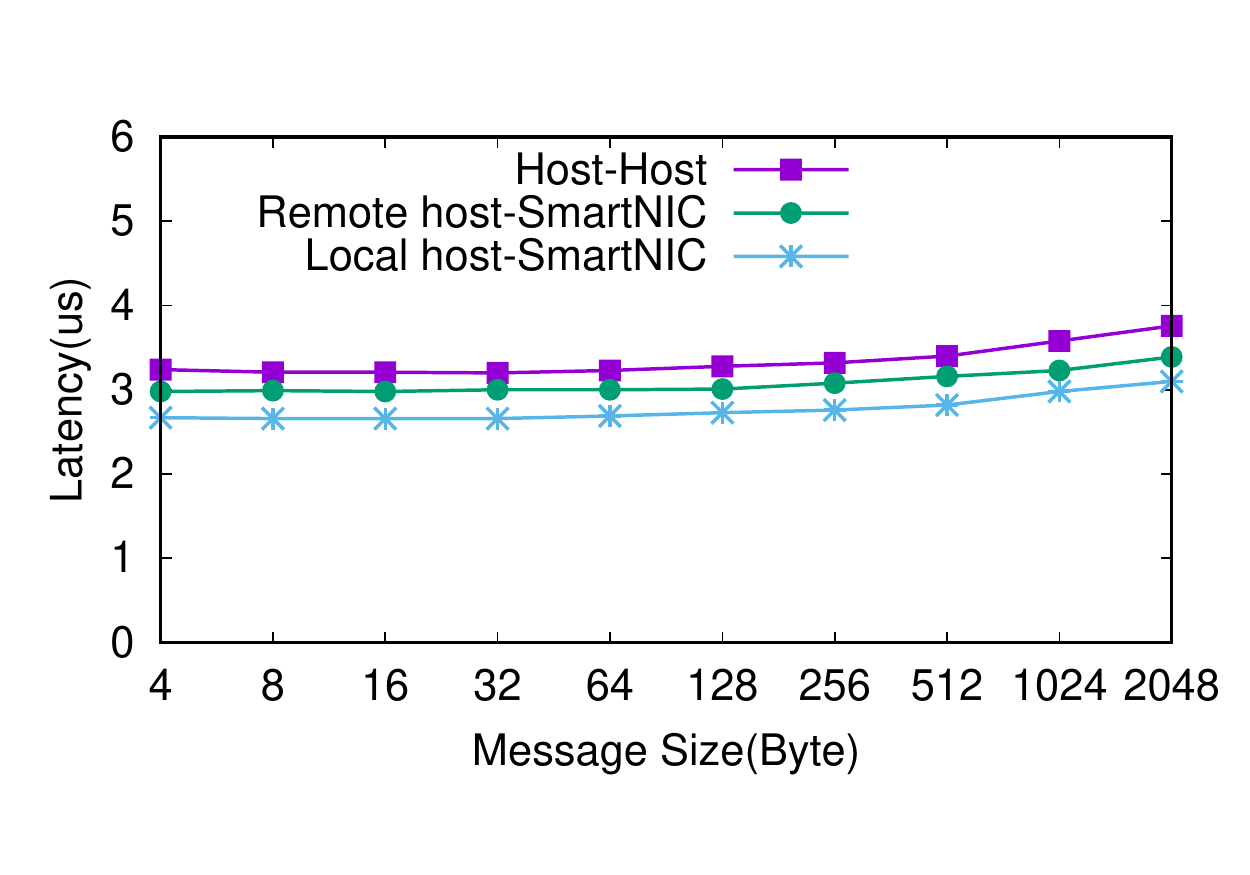}
\end{minipage}%
}%
\subfigure[RDMA Send latency.]{
\begin{minipage}[t]{0.32\linewidth}
\centering
\includegraphics[width=2.4in,trim=0 30 0 30,clip]{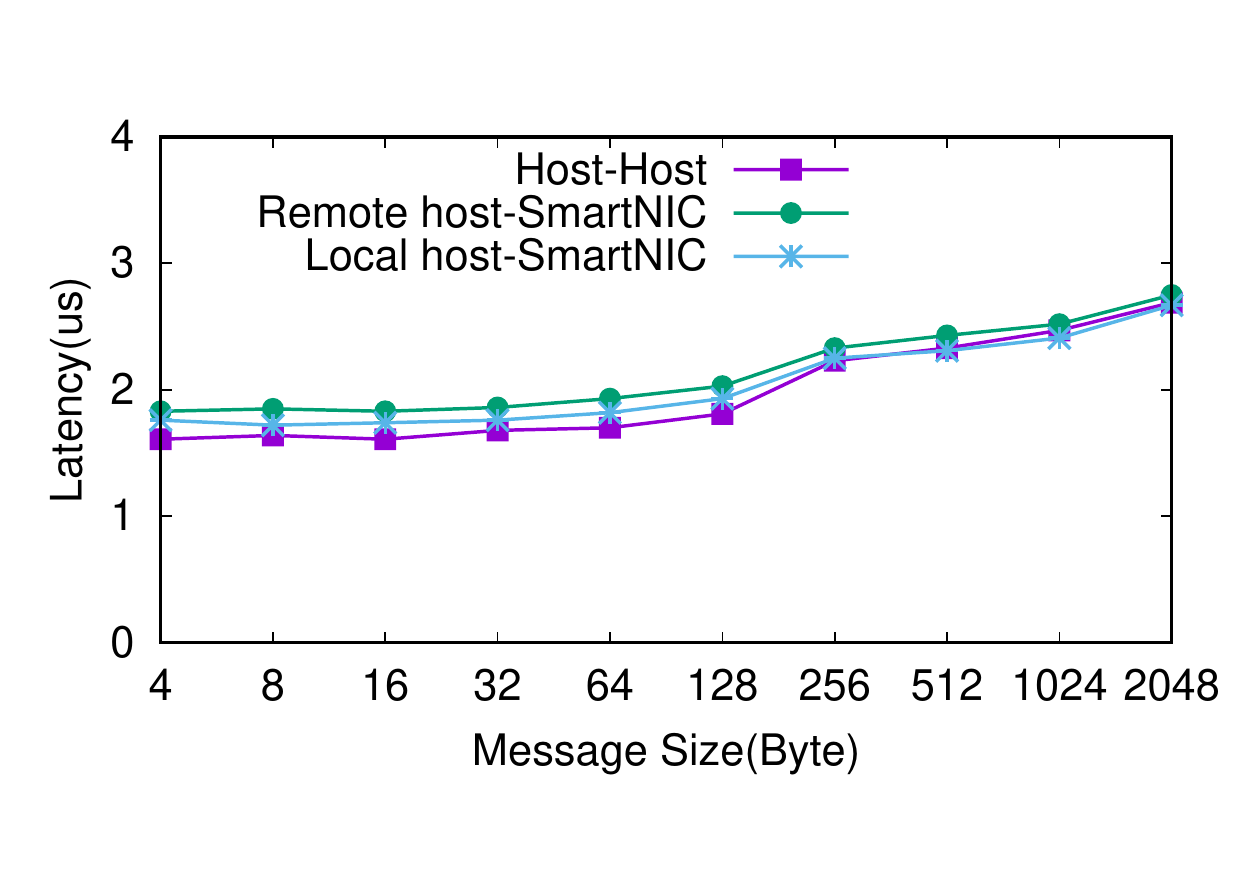}
\end{minipage}
}%

\centering
\caption{The overhead of communication between the SmartNIC and the host.}
\label{figrdma}
\end{figure*}

\subsubsection{Memory Access}
Since the efficiency of executing a task is affected by the latency of memory access, we also measure the performance of the on-board memory on an off-path SmartNIC. We use \texttt{sysbench}\cite{Sysbench} to measure the memory access latency of the SmartNIC and host. Our experiment includes four scenarios: random read, random write, sequential read and sequential write. The block size of the memory access varies from 8B to 4096B. The result is shown in Figure~\ref{figmem}. We find that regardless of random access or sequential access, the memory access latency of the SmartNIC is always higher than that of the host. Especially for random write, the latency of the SmartNIC is much higher than that of the host when the memory block is large. This larger memory access overhead will exacerbate the performance degradation when tasks are offloaded to the SmartNIC.

\subsubsection{Communication Overhead}
In order to fully understand the characteristics of SmartNICs, it is also important to explore the communication overhead between the SmartNIC (here we mean the SoC of the SmartNIC) and the host. Unlike on-path SmartNICs, an off-path SmartNIC does not have low-level interfaces to manipulate host memory\cite{schuh2021xenic}, so it can only communicate with the host through network protocols such as RDMA and TCP. We use the \texttt{perftest}\cite{Perftest} tool to test the end-to-end RDMA latency between two hosts, from the remote host to the local SmartNIC, and from the local host to the local SmartNIC, respectively. Our evaluation includes three RDMA operations: Write, Read, and Send.

Under different payload sizes, the latency of Write and Send operations from the host to the SmartNIC is a little higher or close to the latency between two hosts as shown in Figure~\ref{figrdma}. Only the Read latency from the host to the SmartNIC is a little lower than the latency between two hosts. It is because the communication between the host and the SmartNIC core needs to go through the complete network stack and the NIC switch. This shows that the communication overhead between the SmartNIC and the host cannot be ignored, which is different from on-path SmartNICs. When an off-path SmartNIC communicates with other devices, it acts like a new endpoint in the network. We should try to avoid this part of overhead in the following design.

\subsection{Design Guidelines}
Based on the performance characterization and the SmartNIC structure, we present the following design guidelines.

\textbf{Guideline 1. Accelerate specific applications with dedicated hardware accelerators on the SmartNIC.} In addition to the programmable ARM cores, off-path SmartNICs are usually equipped with special accelerators for offloading applications. For BlueField, the equipped accelerators include a regular expression matching accelerator and encryption/decryption accelerators. If tasks are offloaded to these accelerators, they are executed by special hardware. This is much faster than being executed on general-purpose processors. BlueField provides the DOCA framework, which provides convenient interfaces for developers. Through these interfaces, developers can directly use the underlying hardware accelerator to offload tasks from the host to the SmartNIC. This approach can achieve better performance than host-based implementation. In addition, it requires only a little development overhead and it is easy to use. However, these accelerators lack programmability and cannot meet the needs of customization. Therefore, in this respect, it is inferior to FPGA-based SmartNICs, which can consolidate arbitrary logic.

\textbf{Guideline 2. Offload latency-insensitive background operations to the SmartNICs.} The previous experimental results tell us that the general-purpose processor cores of off-path SmartNICs are much weaker than the host cores, so directly moving the program executed on the host to the SmartNIC for execution will cause performance degradation. This has been reported in previous work\cite{kim2021linefs}. At the same time, the slower memory access and huge communication overhead between the SmartNIC and the host will aggravate this situation. In order to avoid these shortcomings of off-path SmartNICs, we suggest selecting latency-insensitive background operations to be offloaded to SmartNICs. Background operations are common in many systems. They are often separated from front-end operations such as responding to client requests. For example, in a file system, some operations in log processing are background operations. These operations have no explicit latency requirements, but consume a large amount of host CPU resources. Such a background operation can be decoupled from the front-end operations easily, so it would be a suitable object to offload. Although the SmartNIC takes longer time to perform these operations than the host, this design does not slow down the front-end interaction with the client because the front-end service process does not remain blocked until the background processing is completed. Offloading such latency-insensitive operations can reduce the load on the host and save host CPU cycles. The saved CPU cycles can be used to respond to clients or perform other latency-sensitive tasks, thereby improving overall performance.

\textbf{Guideline 3. Treat the SmartNIC as a new endpoint in the network to expand the computing power and storage space of hosts.} Off-path SmartNICs have highly programmable processors, DRAM, and non-volatile storage media. For example, BlueField has 8 ARM cores, 16GB DRAM, and eMMC flash memory, while our host has 32 Intel cores and 64GB DRAM. Although SmartNICs have smaller computing and storage resources than hosts, they are still sufficient to undertake a certain amount of workload. In addition, this type of SmartNIC has complete network stacks such as RDMA and TCP and it has an independent IP address that is different from the host. So it can communicate with the local host and the remote host easily. It is also important that an off-path SmartNIC has a complete operating system, so many applications on the host can also run on the SmartNIC. As a result, the SmartNIC can be used as an independent node in the network, as a horizontal expansion of server host resources. For example, we can let the host and the SmartNIC execute the same task simultaneously, which improves the parallelism and throughput of the system. These are things that on-path SmartNICs cannot do.

\textbf{Guideline 4. Avoid directly employing the design method of systems based on on-path SmartNICs.} Most of the current research on multi-core SoC SmartNICs is based on on-path SmartNICs, but it is not feasible to directly use their methods to offload tasks to off-path SmartNICs. This is because the structure and performance of the off-path SmartNICs are much different from theirs. The processor cores of the on-path SmartNICs are weaker than host cores and cannot perform complex operations, but the overhead between them and the hosts is much lower. Therefore, Xenic\cite{schuh2021xenic} uses an on-path SmartNIC as a cache to store the most frequently accessed data to reduce the PCIe overhead of accessing host memory. If an off-path SmartNIC adopts such a scheme, the huge communication overhead between the NIC and the host memory will make the performance worse when cache miss appears. Other works\cite{qiu2021automated,Floem} manipulate incoming and outgoing data packets by leveraging low-level interfaces provided by on-path SmartNICs to improve the performance of network functions. However, off-path SmartNICs do not have such interfaces, so they cannot process the data packets easily. Therefore, we should not directly employ the design methods for on-path SmartNICs. 

Among the above guidelines, \textbf{Guideline 1} suggests directly using the accelerators on the SmartNIC to achieve better performance. \textbf{Guideline 2} and \textbf{Guideline 3} show how to use the general-purpose ARM cores on the SmartNIC to improve existing systems. \textbf{Guideline 4} advises developers to avoid some pitfalls of using off-path SmartNICs.

\section{Case Study}
\label{seccase}
In this section, we demonstrate how the design guidelines work in practice. We apply these guidelines to some commonly used applications. Then we show the feasibility of these guidelines through a series of experiments. Section \ref{secg1} to Section \ref{secg4} are corresponding to the 4 guidelines.

\subsection{Using Accelerators}
\label{secg1}
Off-path SmartNICs like BlueField are equipped with many dedicated accelerators as described in Section \ref{sec2}. Operations that can take advantage of the accelerators can perform better on the SmartNIC than on the host, so we consider offloading these operations from the host to the SmartNIC and using the accelerator to handle them. Since BlueField is equipped with an accelerator called regular expression processor (RXP), we implement a regular expression matching application. We use the interfaces provided by the DOCA framework to call the accelerator.

Regular expression matching can be used in many scenarios, including detecting incoming and outgoing traffic, filtering threatening packets, and web log analysis. These operations cannot be executed efficiently on the host compared with special hardware. Therefore, it is a good choice to use the SmartNIC to offload the execution of these tasks.

In our regular expression matching application, all regular expressions are stored in a text file called rule file. First, the rule file is compiled by RXP compiler (RXPC) to generate an RXP object format (ROF) binary file. Then, we initialize the context of regular matching, including finding the SmartNIC based on the PCIe address, setting the number of required queue pairs (QPs), and allocating memory space for each QP. Each QP contains an input queue and an output queue. The accelerator can get data from the input queue and matching results are written to the output queue. After that, we load the compiled ROF file as matching rules and start the accelerator. Now the initialization completes. We divide the scan of the entire log into multiple jobs. Then the program enters a loop until the number of bytes to be processed becomes zero. The program continuously enqueues pending jobs into the input queue of the accelerator QP, dequeues completed jobs from the output queue and records the matching results, including a list of offset and matched length.

\begin{table}[htbp] 
\caption{Throughput (Gb/s) of regular expression matching.}
\label{regex}
\centering
\begin{tabular}{lll}
\hline
    Performance & SmartNIC & Host \\
\hline
    Overall throughput & 30.87 & 27.74 \\
    Maximum throughput & 32.12 & 28.82 \\
    
\hline   
\end{tabular}
\end{table}

\begin{figure*}[!t]
\centering
\subfigure[Throughput.]{
\begin{minipage}[t]{0.32\linewidth}
\centering
\includegraphics[width=2.4in,trim=0 30 0 30,clip]{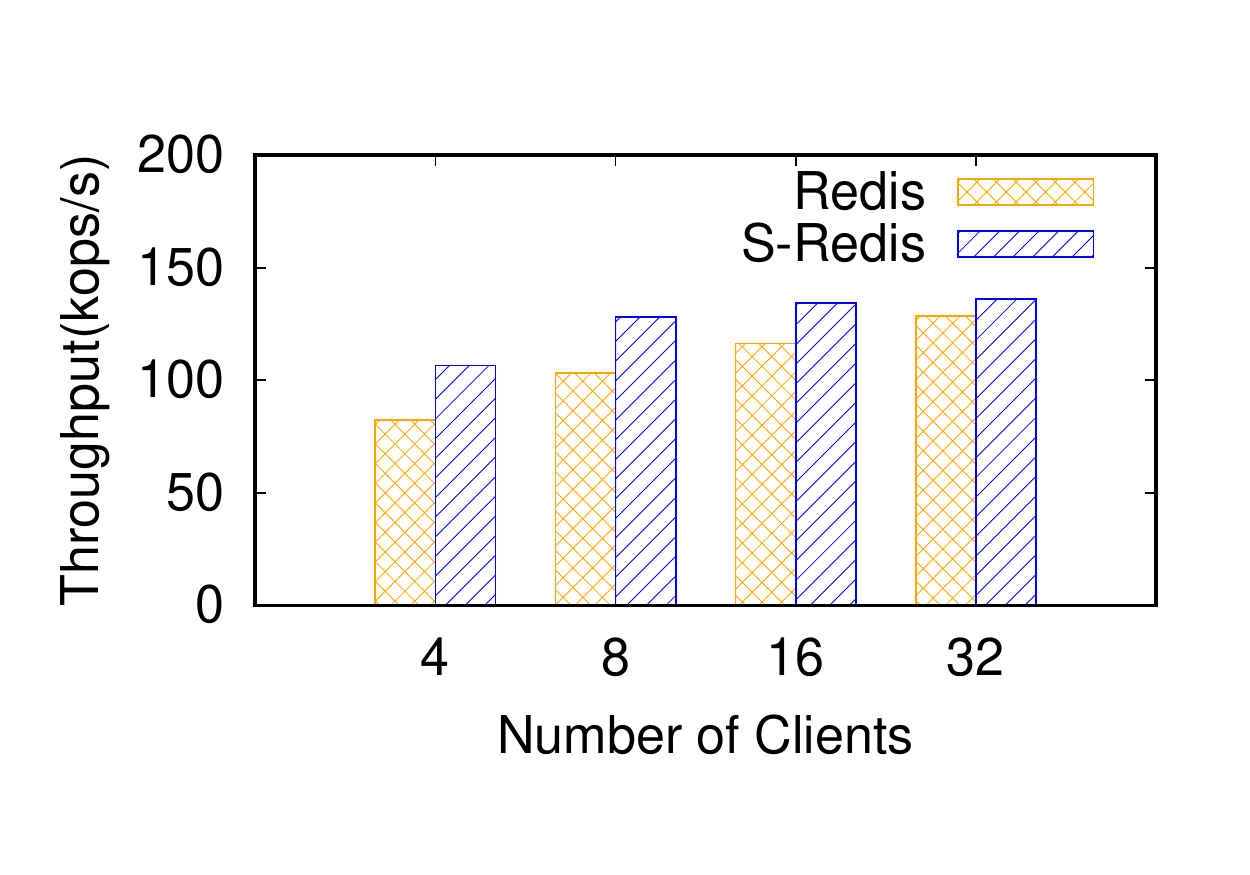}
\end{minipage}%
}%
\subfigure[Average latency.]{
\begin{minipage}[t]{0.32\linewidth}
\centering
\includegraphics[width=2.4in,trim=0 30 0 30,clip]{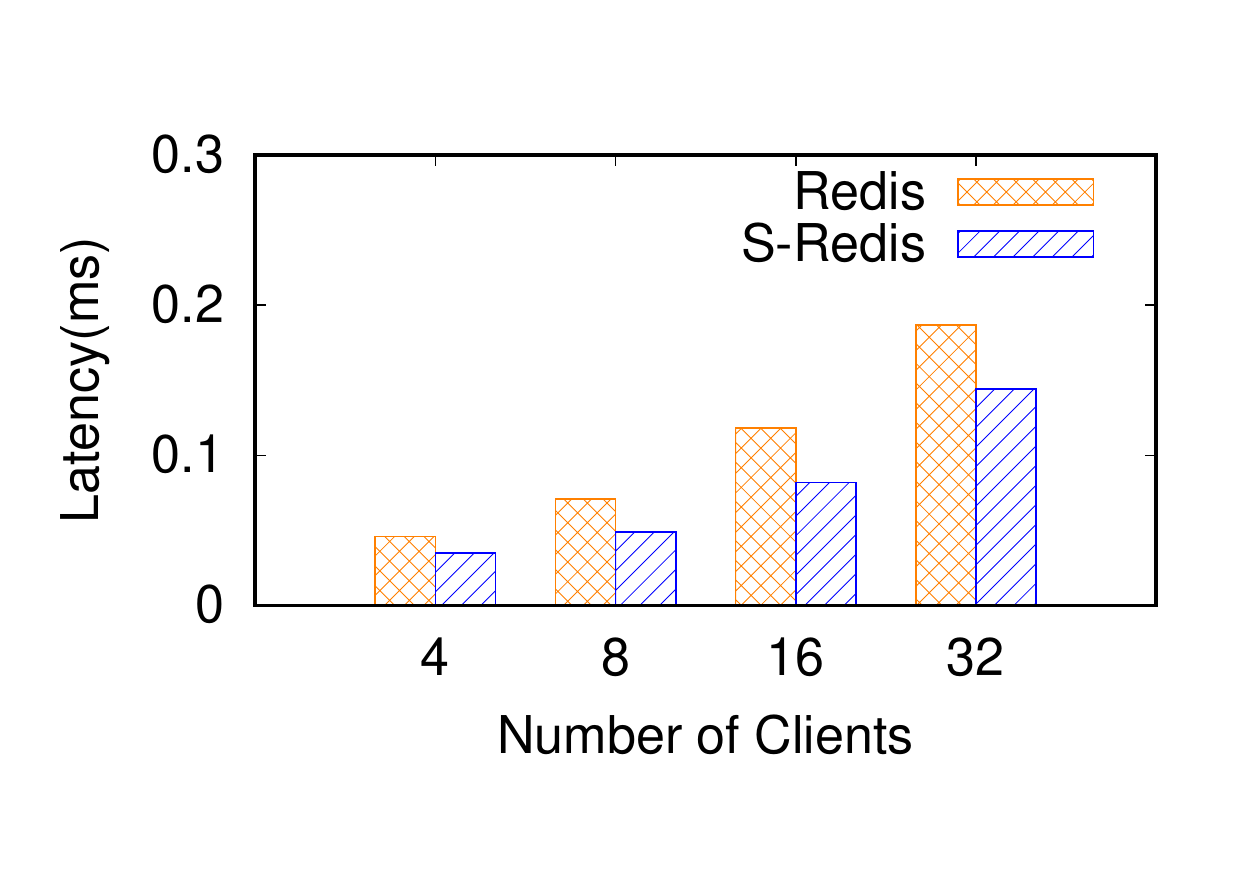}
\end{minipage}%
}%
\subfigure[P99 tail latency.]{
\begin{minipage}[t]{0.32\linewidth}
\centering
\includegraphics[width=2.4in,trim=0 30 0 30,clip]{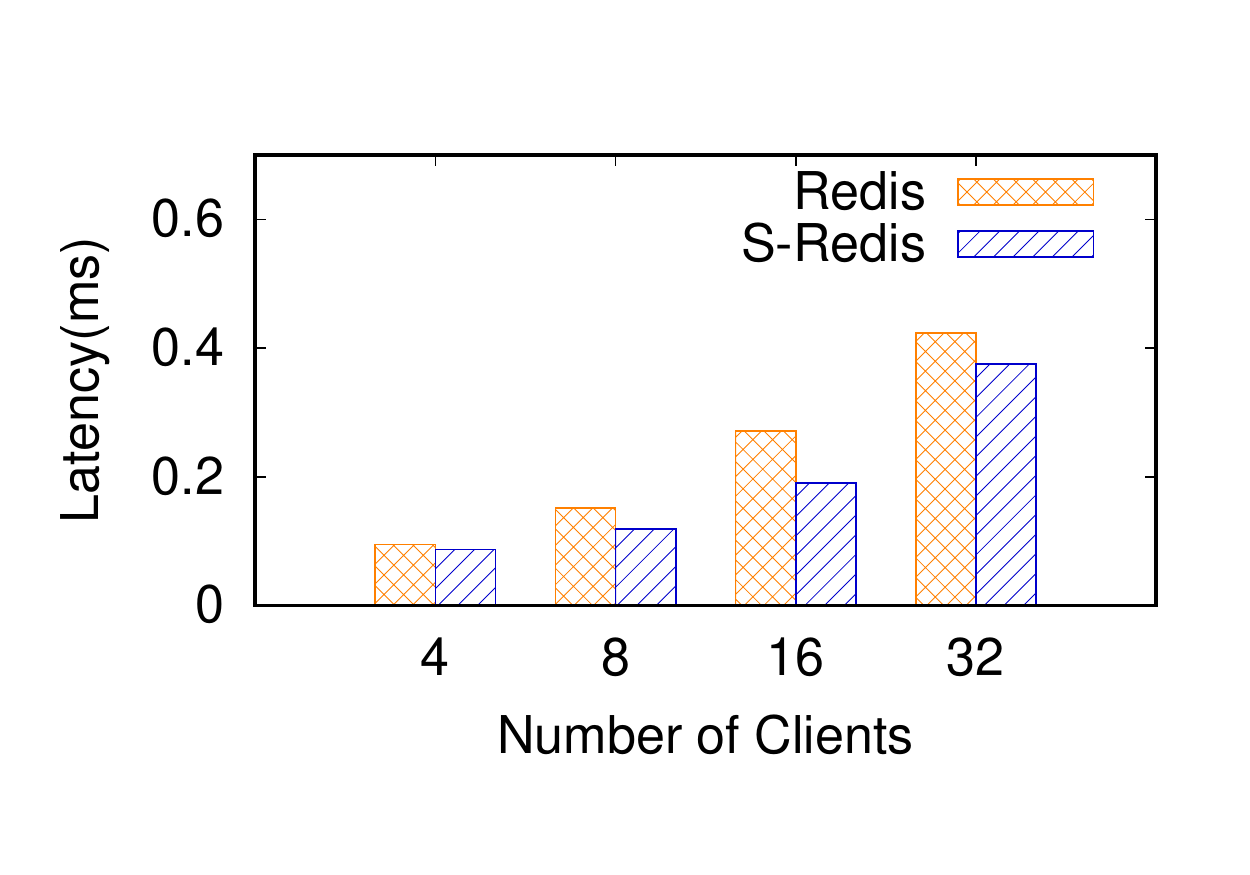}
\end{minipage}
}%

\centering
\caption{The performance of S-Redis when executing SET commands compared with original Redis when there are 3 slaves.}
\label{fig3c}
\end{figure*}

We use a web log dataset to evaluate the performance of regular expression matching. Web log analysis includes collecting statistics on the distribution of accessed pages and user IP addresses. We compare the performance of offloading the matching of regular expression to the SmartNIC (SmartNIC) with the performance of matching using the Hyperscan library on the host (Host). The experimental results are shown in Table \ref{regex}. Offloading this task to the accelerator on SmartNIC can achieve an overall throughput of 30.87Gb/s, which exceeds the throughput of the host by about 11\%. In addition, the maximum throughput is improved by about 12\% compared with the baseline. So using the accelerator to offload can get better performance compared with the host-based method. The general-purpose Arm cores on the SmartNIC can also use the Hyperscan library to match expressions, but this will make the performance worse as the SmartNIC cores are much weaker than the host.

However, this method has limitations. The DOCA framework only provides high-level encapsulations, which makes it difficult to meet different needs of various systems. Moreover, as the hardware accelerator used here is not programmable, using it to develop applications is not very flexible and it is hard to offload other complicated tasks.

\subsection{Offload Background Operations}
Due to the weaker computing power and larger communication overhead of the SmartNIC, we suggest offloading latency-insensitive background operations in \textbf{Guideline 2}. We use this method to improve the performance of a distributed key-value store. We take Redis as an example like SKV\cite{sun2022skv}.

Redis\cite{Redis} is a popular high-performance key-value database. Redis stores data in memory, providing low latency and high throughput. Moreover, Redis has a distributed mode. In this mode, Redis performs data backup in master-slave mode for higher availability and scalability. Redis nodes are divided into a master node and several slave nodes. The master node can accept both write and read requests, but the slave nodes can only be used for reading. After a connection is established between nodes, every time the master node receives a write command, it will replicate it to all slave nodes to ensure data consistency. The slave node executes this command as soon as it receives it. Note that the master will not remain blocked until the slaves acknowledge that the replication has finished unless the client sends a WAIT command. During replication, the master node can normally execute received commands and return results to clients as soon as possible. Therefore, as a background operation in the key-value store, replication does not have strict latency requirements and it is a suitable object for offloading. We adopt Redis as a use case here because Redis is widely used now.

We offload the replication operation to the SmartNIC as SKV does. To optimize the performance of the key-value store, a SmartNIC is installed on the master node. The SmartNIC does not store key-value data. All the data is stored on the host. The SmartNIC is only used to replicate data from the master node to slave nodes. Unlike SKV, we adopt the design of the network communication module in the original Redis. In order to offload the replication operation, a replication list is stored on the SmartNIC, which contains the information of the master node and all slave nodes. At initialization, all slave nodes send their addresses and ports to the SmartNIC. The SmartNIC records this information in the replication list.


\begin{figure}[htbp]
\begin{center}
\includegraphics[width=2.8in]{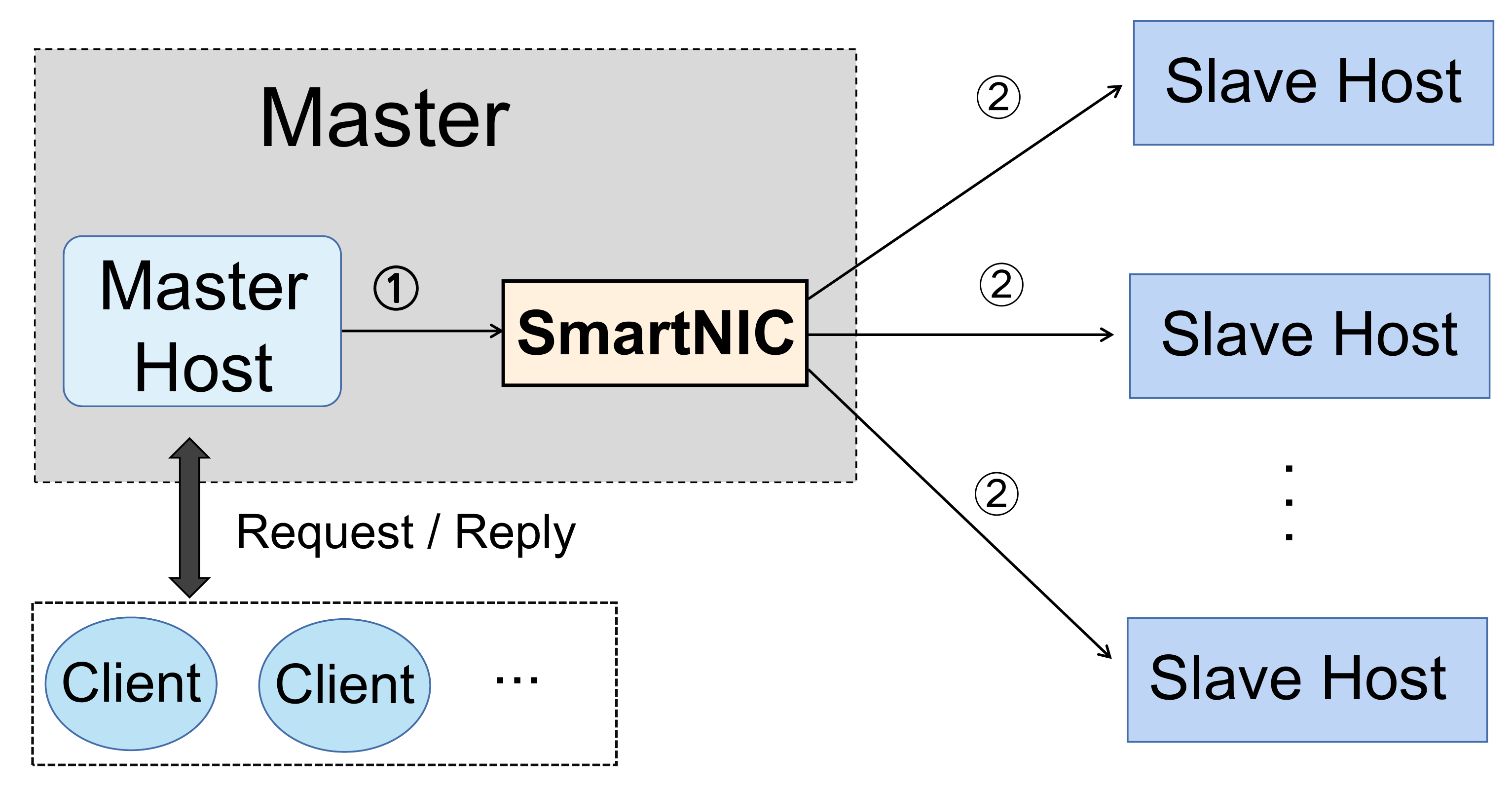}
\caption{The process of using a SmartNIC to offload replication.}
\label{g2}
\end{center}
\end{figure}

When the master node receives a Redis command from the client, it first checks whether the command will change the data in the database. If not, the master node will execute the command as normal. If it does, for example it is a SET command or a DEL command, the master node needs to replicate the command after execution to ensure data consistency between nodes. The process of replication is shown in Figure~\ref{g2}. The master host sends a message to the SmartNIC, which includes the command to replicate \ding{192}. After the SmartNIC receives it, it replicates the command to all slave nodes according to the replication list \ding{193}. Every slave node executes the command immediately after receiving it. Now the replication of a command between the master and slave nodes is completed.

We compare the throughput and latency of Redis accelerated with SmartNIC (S-Redis) with the original Redis. The master-slave relationship between nodes has been established before the experiment. SET commands are sent to the master node for execution. In this experiment, we first test the case of three slave nodes and use \texttt{redis-benchmark} to evaluate the performance, the same as SKV.

As shown in Figure \ref{fig3c}, the throughput and latency of S-Redis are better than the original Redis when the number of concurrent clients varies. Under 8 concurrent connections, the throughput of S-Redis is 24\% higher than the original Redis, while the average latency is reduced by 31\%, and the tail latency is reduced by 22\%. This is because in Redis replication mechanism, the master node server needs to send each SET command received from the client to all the slave nodes one by one. Each time it is sent, it consumes host CPU cycles. In S-Redis, the master node only needs to send one message to the SmartNIC, which reduces the CPU consumption in data transmission. Due to the offloading design, the replication operation is performed by the SmartNIC. In this way, the saved CPU cycles can be used to send responses to clients, allowing the master server to respond to more client requests in the same period of time. Moreover, many distributed systems have replication mechanisms similar to Redis, so this design can also potentially benefit these systems.

We also discover that the performance improvements brought by SmartNIC offloading in our experiments are greater than the results reported in SKV. In SKV, offloading with SmartNIC only improves the throughput by about 14\%, but here we improve the throughput by 24\%. The performance of SKV is compared to RDMA-based Redis and both of them use RDMA for network communication. However, the original Redis and S-Redis we design use TCP as the network layer. In data replication, TCP data transmission involves switching between user mode and kernel mode and it causes multiple memory copies. Using the SmartNIC to reduce this part of the overhead can reduce more CPU load, resulting in more obvious performance improvement.

\begin{figure*}[!t]
\centering
\subfigure[Throughput.]{
\begin{minipage}[t]{0.32\linewidth}
\centering
\includegraphics[width=2.4in,trim=0 30 0 30,clip]{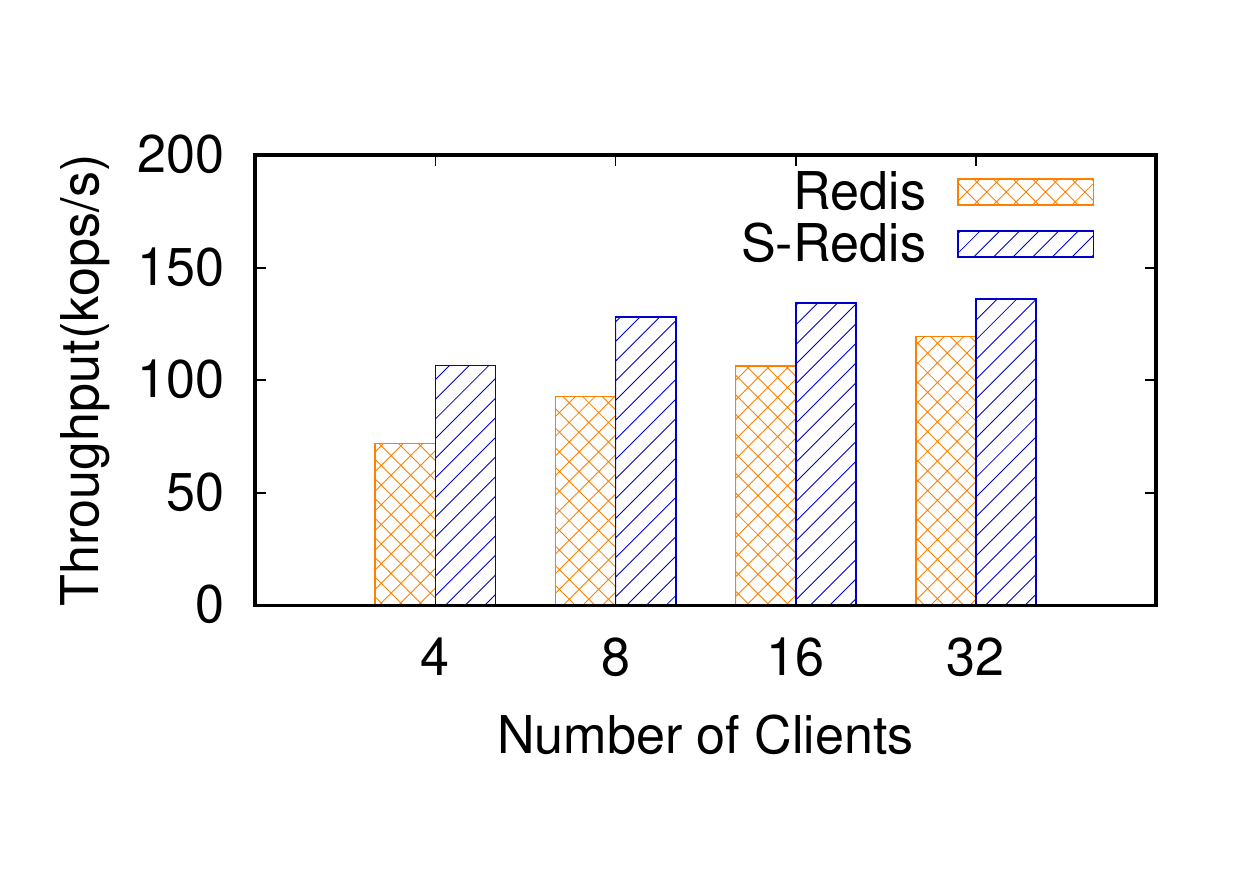}
\end{minipage}%
}%
\subfigure[Average latency.]{
\begin{minipage}[t]{0.32\linewidth}
\centering
\includegraphics[width=2.4in,trim=0 30 0 30,clip]{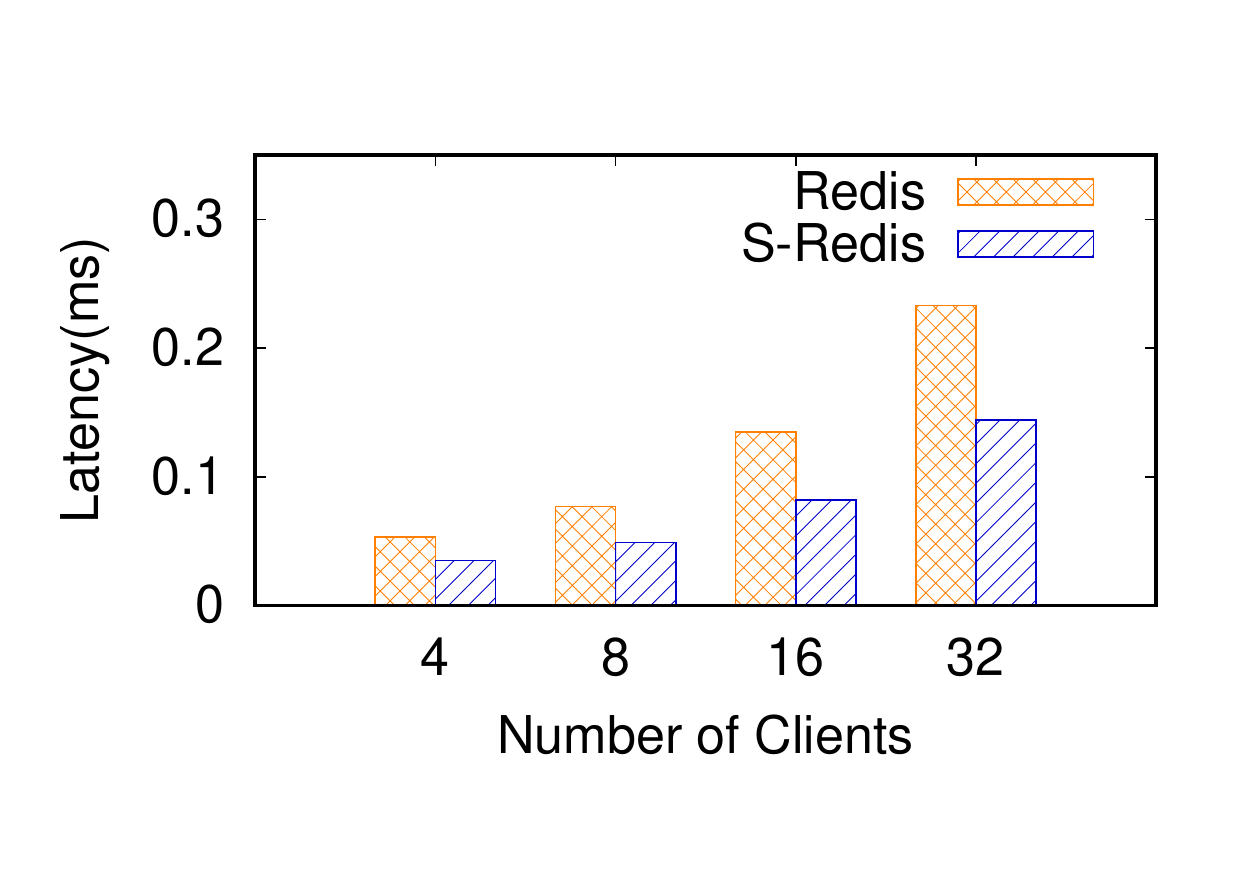}
\end{minipage}%
}%
\subfigure[P99 tail latency.]{
\begin{minipage}[t]{0.32\linewidth}
\centering
\includegraphics[width=2.4in,trim=0 30 0 30,clip]{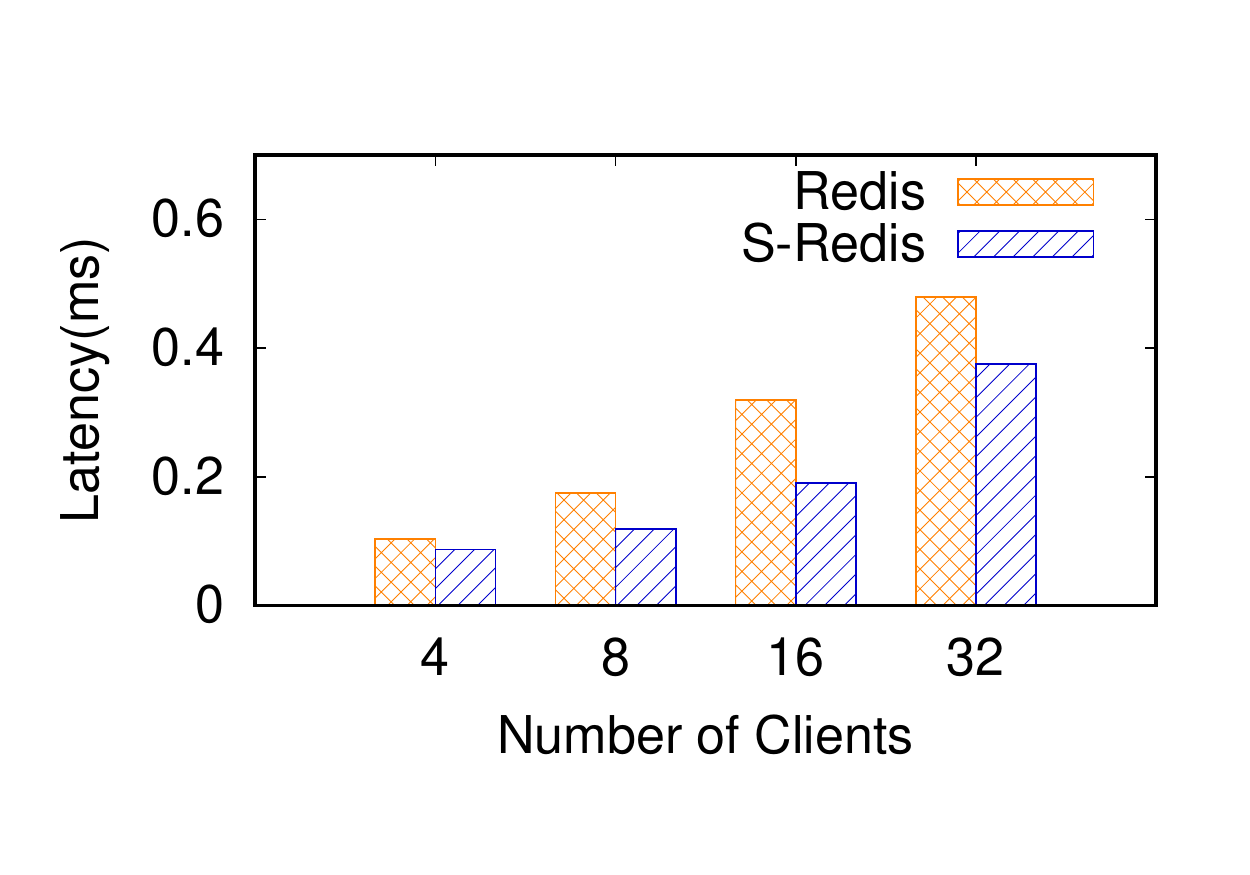}
\end{minipage}
}%

\centering
\caption{The performance of S-Redis when executing SET commands compared with original Redis when there are 5 slaves.}
\label{fig5c}
\end{figure*}

In addition, we test the situation of 5 slave nodes. As is shown in Figure \ref{fig5c}, we find that in this case offloading has a larger performance improvement, both in latency and throughput. The throughput of S-Redis is 39\% higher than the original Redis, while the average latency is reduced by 37\% and the tail latency is reduced by 32\%. With more slave nodes, the overhead of replication is larger. This result shows that using the SmartNIC to offload heavier latency-insensitive background operations can bring more significant performance improvements.

\subsection{Treat the SmartNIC as a New Endpoint}
An off-path SmartNIC has a complete operating system, which enables programs that can run on a general host to be executed on the SmartNIC. This is the biggest difference from on-path SmartNICs. In addition, the BlueField SmartNIC has 16GB on-board memory. Although it is smaller than our host memory which is 64GB, it still can undertake a certain amount of workload. Due to this feature, \textbf{Guideline 3} proposes that SmartNICs can do the same thing as the host to expand host resources.

To show how to apply this guideline, here we take Redis as an example to implement data sharding on the host and the SmartNIC. The purpose of sharding is to expand the capacity of data storage, because the storage space of a host is limited. At the same time, distributing data to multiple nodes enables multiple clients to access data in different shards simultaneously, improving system concurrency and overall throughput. In production deployments of Redis, this pattern is widely used and it is called horizontal scaling. Therefore, here we regard the SmartNIC as an independent endpoint and let it be a node of data shard.

\begin{figure}[htbp]
\begin{center}
\includegraphics[width=2.8in,trim=100 100 100 100,clip]{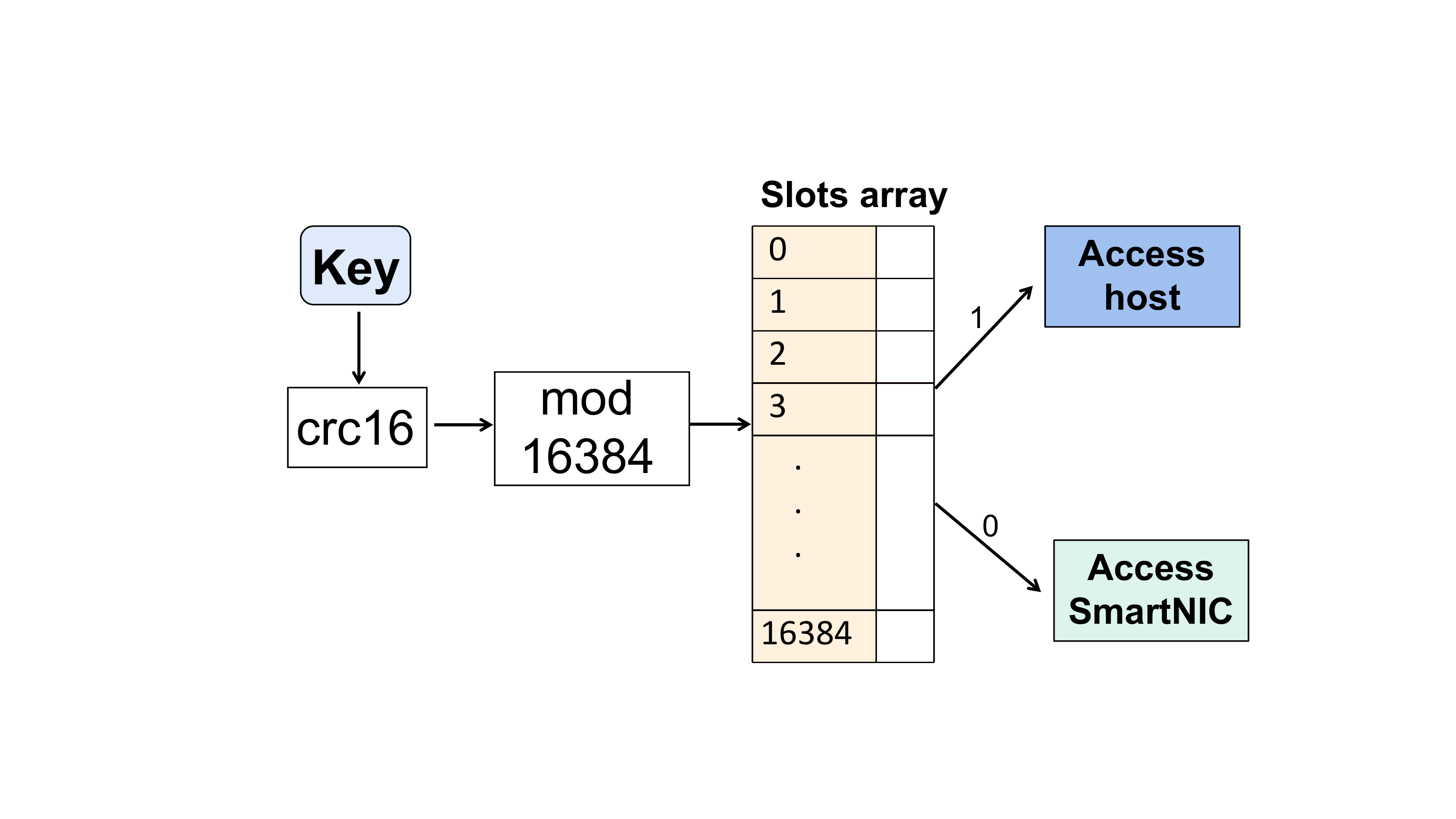}
\caption{The process of finding the location of a key.}
\label{g3}
\end{center}
\end{figure}


Our host and SmartNIC store non-overlapping data shards. A simple sharding method is range sharding, where objects in a certain range are mapped to the same node. However, the disadvantage of range sharding is that when the workload is full of sequential writes, there will be hot spots. In addition, range sharding needs to store multiple ranges in a table, which requires complex management. This method is relatively inefficient. Instead, hash sharding is adopted here. The key space is divided into 16384 hash slots. When the client accesses the data, it first calculates a 16-bit value for the key according to the CRC16 algorithm, and then modulo it with 16384 as shown in Figure \ref{g3}. In this way, it is easy to figure out which hash slot the key resides in. Each hash slot can store multiple keys, and each key belongs to a hash slot. The client then queries the Slots array, which is an array of binary bits. It is 2048 bytes long and contains 16384 binary bits, corresponding to 16384 hash slots. If the bit in the slots array is 1, it means that the hash slot is on the host. On the contrary, if it is 0, it means that the hash slot is located on the SmartNIC. Then the client can find the location of the key to access. Since the time complexity of this operation is O(1), the overhead of this part can be ignored.

\begin{figure}
\begin{minipage}[t]{0.47\linewidth}
\includegraphics[width=\linewidth,trim=20 0 20 0,clip]{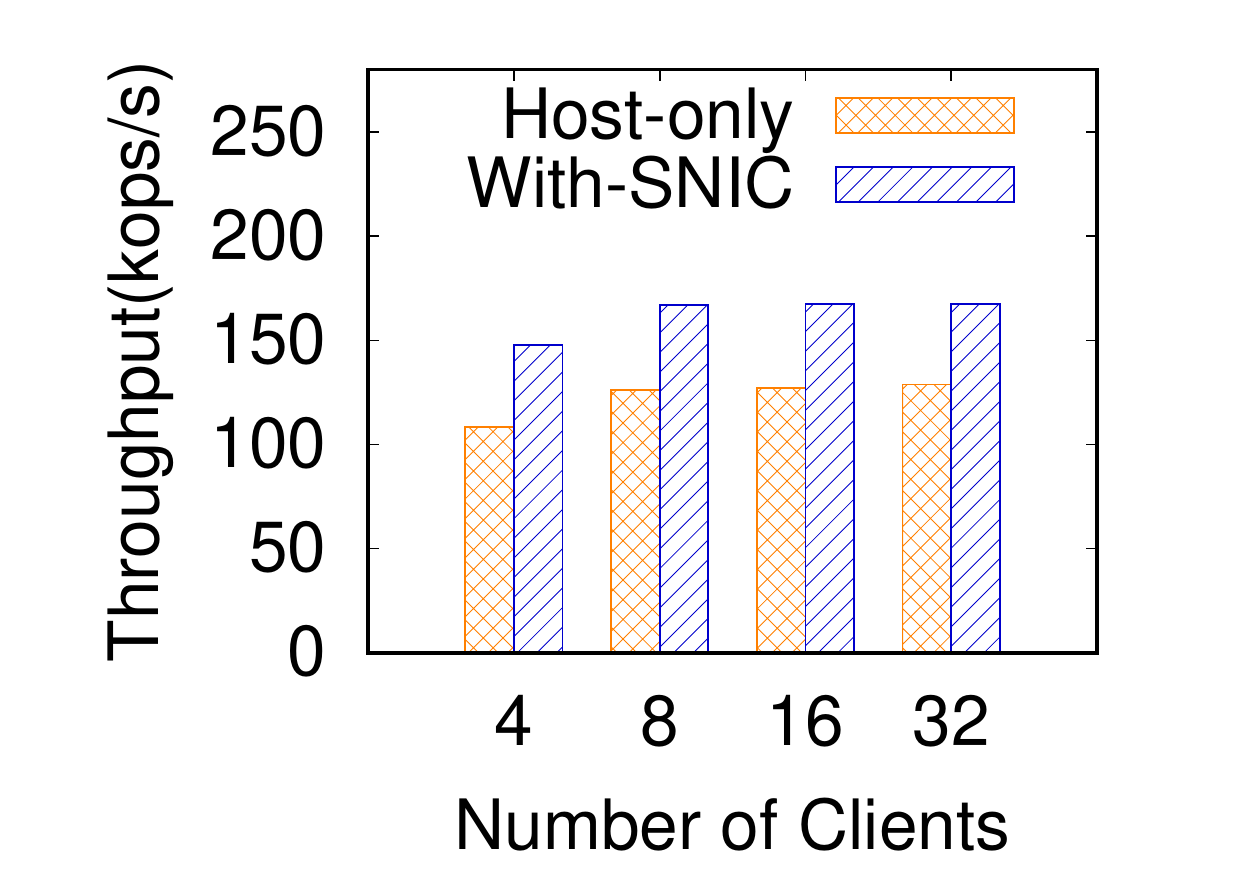}
\caption{The performance of Redis data sharding under different number of concurrent clients.}
\label{g3exp1}
\end{minipage}%
\hfill%
\begin{minipage}[t]{0.47\linewidth}
\includegraphics[width=\linewidth,trim=20 0 20 0,clip]{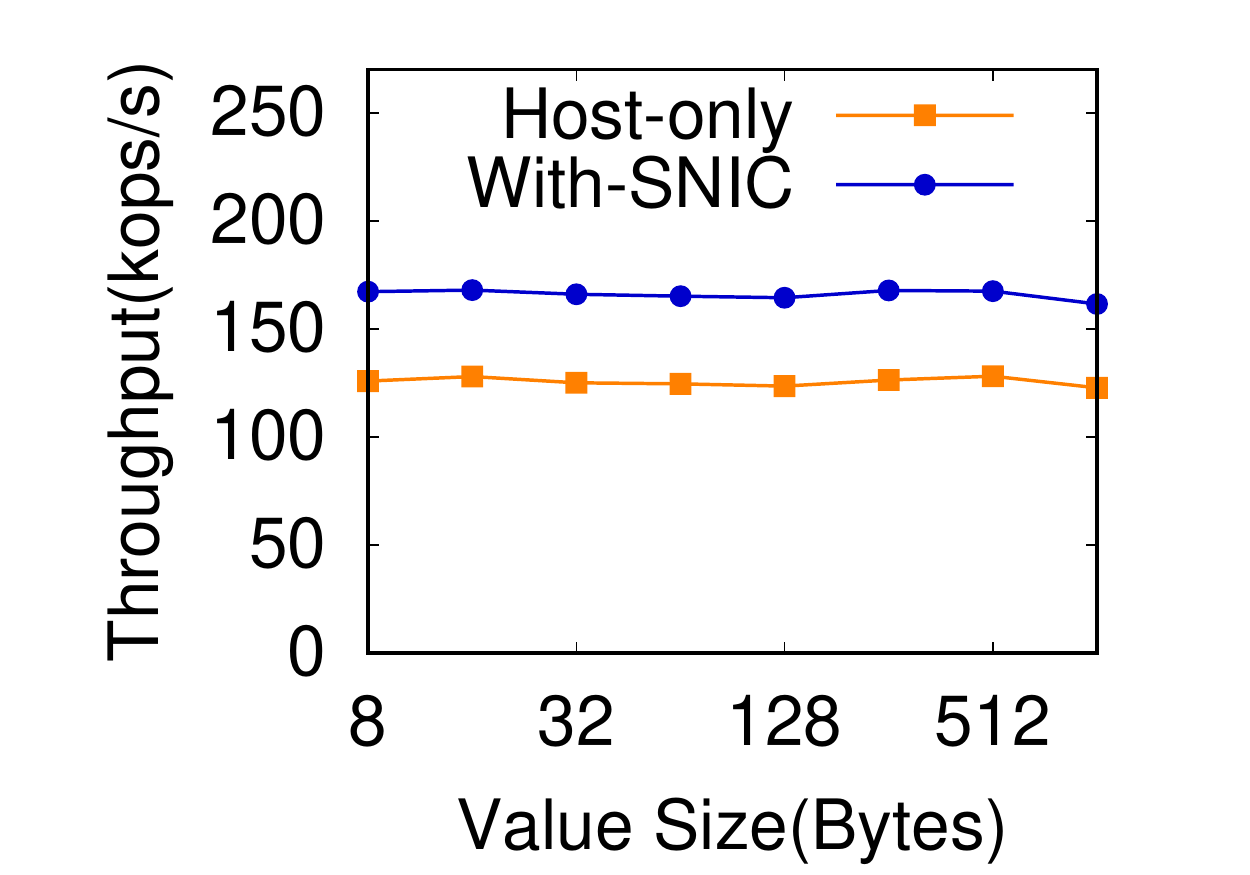}
\caption{The performance of Redis data sharding under different value sizes.}
\label{g3exp2}
\end{minipage}
\end{figure}

We test the performance under different number of client connections. We evaluate the performance of executing the SET command of With-SNIC (the SmartNIC and the host store different data shards) and Host-only (only the host stores data). Since the host and the SmartNIC can process requests in parallel, the throughput of With-SNIC is 30\% higher than that of Host-only as shown in Figure~\ref{g3exp1}. Then we test the throughput under different value sizes. The value size increases from 8B to 1024B and the result is shown in Figure~\ref{g3exp2}. Under different value sizes, the throughput of With-SNIC is stable, and it is also much higher than the case of Host-only. In the Host-only case, all requests are processed by the host and the SmartNIC remains idle. For With-SNIC, both the host and the SmartNIC are processing requests at the same time and both of them are trying their best to handle requests. This tells us that although the processor core performance of the off-path SmartNIC is not as good as that of the host core, its processing power can still help improve the overall throughput.

Moreover, we apply the idea of \textbf{Guideline 3} to MongoDB. MongoDB is a document-based storage system that is also widely used in industry. The structure of document storage systems is similar to key-value stores, but it is more efficient for querying large-scale data. Such a system stores semi-structured documents and encodes data in formats such as JSON. We use a method similar to the description above to let the SmartNIC and the host store non-overlapping data shards. This enables the SmartNIC and the host to handle concurrent requests at the same time, increasing the overall throughput of document access.
\begin{table}[htbp] 
\caption{YCSB workload}
\label{ycsb}
\centering
\begin{tabular}{lll}
\hline
    Workload & Read-Write-Scan(\%) & Workload Type \\
\hline
    A & 50-50-0 & Write Intensive \\
    B & 95-5-0 & Read Intensive \\
    C & 100-0-0 & Scan Only \\
    D & 95-5-0 & Read Latest \\
    E & 0-5-95 & Short Range Scan \\
\hline   
\end{tabular}
\end{table}

\begin{figure}
\begin{minipage}[t]{0.47\linewidth}
\includegraphics[width=\linewidth,trim=20 0 20 0,clip]{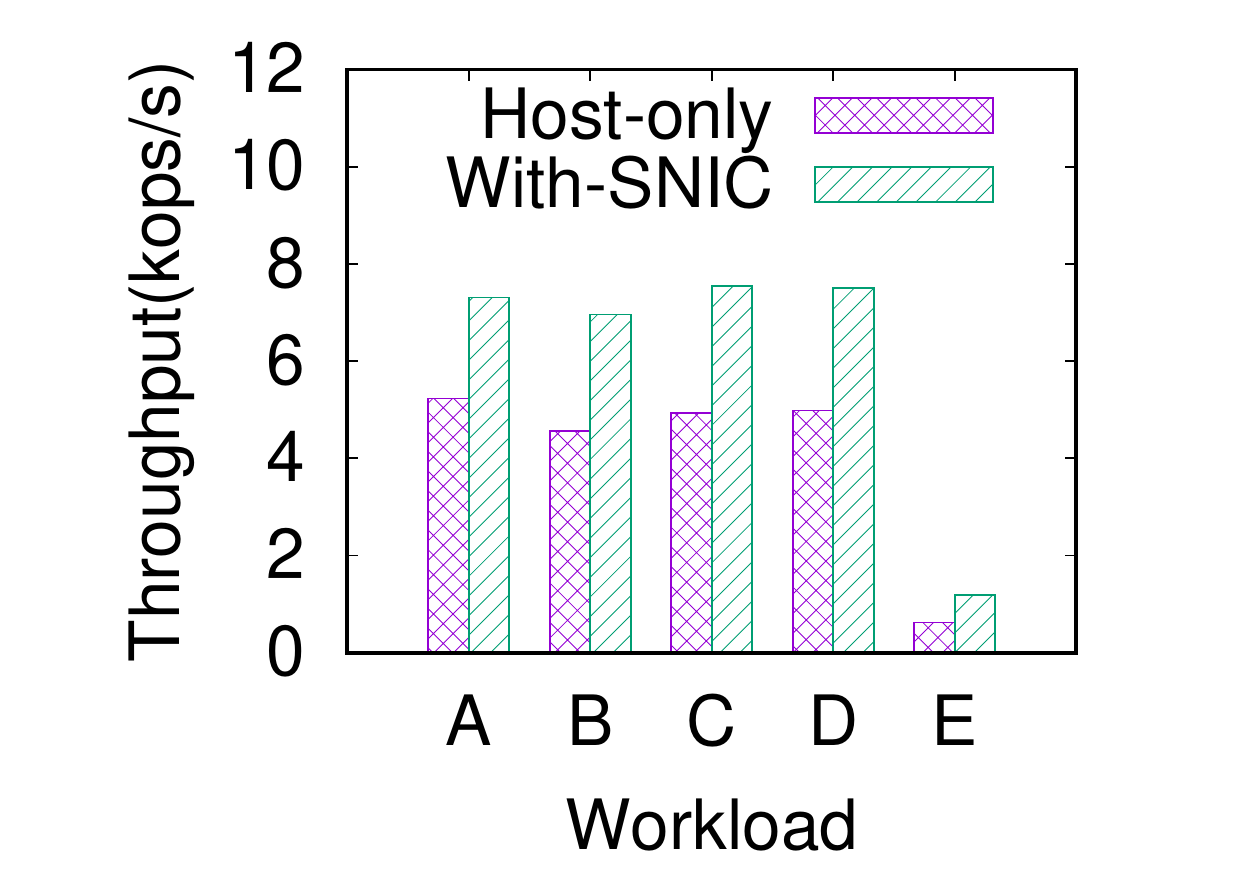}
\caption{The performance of MongoDB data sharding under different YCSB workloads.}
\label{mongo1}
\end{minipage}%
\hfill%
\begin{minipage}[t]{0.47\linewidth}
\includegraphics[width=\linewidth,trim=20 0 20 0,clip]{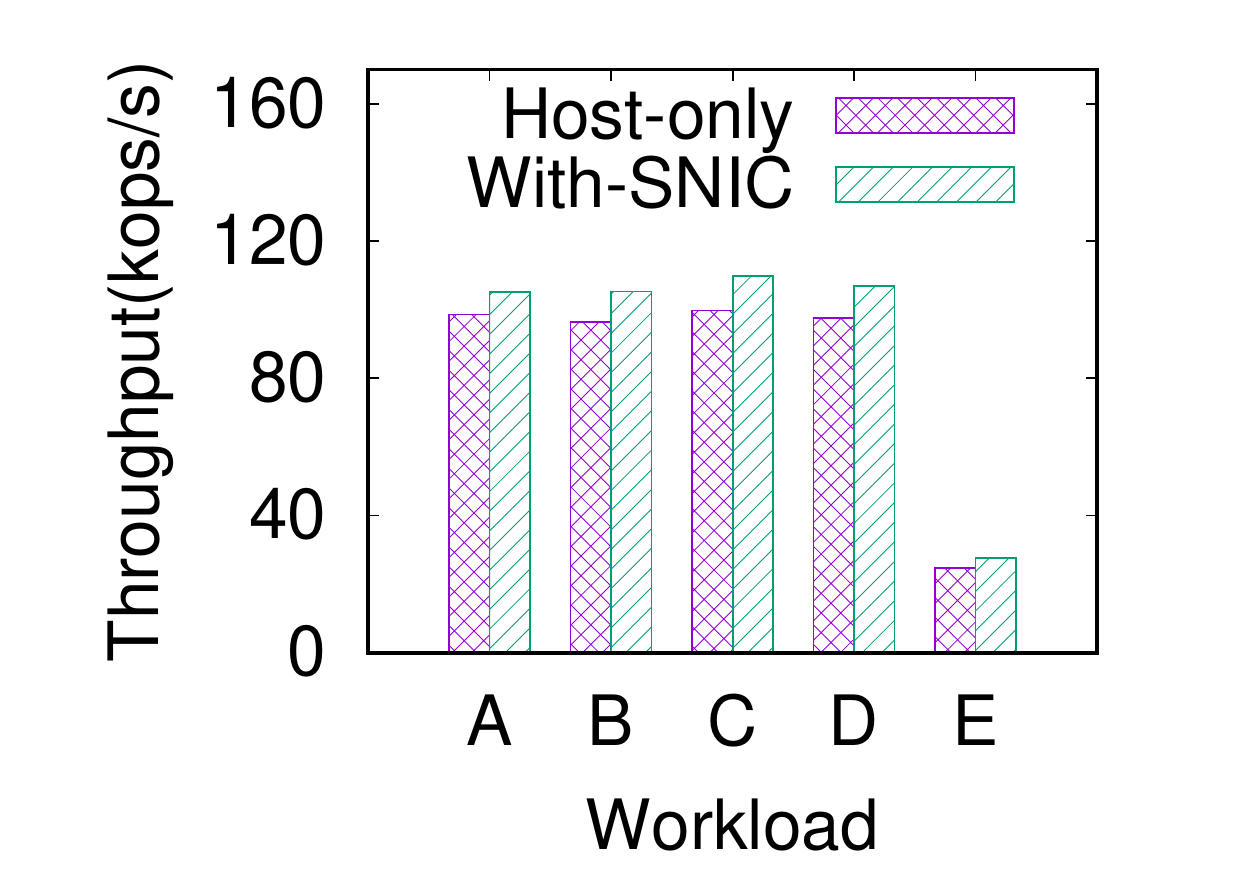}
\caption{The performance of MongoDB data sharding in multi-threading mode.}
\label{mongo2}
\end{minipage}
\end{figure}

We use YCSB\cite{cooper2010benchmarking} benchmark to evaluate the performance improvement. YCSB has different types of workloads as listed in Table~\ref{ycsb}. First, we test the performance of With-SNIC (the SmartNIC and the host store different MongoDB data shards) and Host-only (only the host stores data) in the case of single thread. Under different workloads, With-SNIC outperforms host-only as shown in Figure \ref{mongo1}. The throughput increases by more than 30\% after applying \textbf{Guideline 3}. However, the throughput of MongoDB in the case of single thread is far from the maximum throughput it can reach, which is different from Redis. This is because MongoDB and Redis have different thread models. MongoDB uses a single Listener thread to accept all client connections. Each time the Listener thread receives a new connection, it assigns a new thread to that connection if multi-threading is enabled. This new thread is only responsible for handling requests from this connection. Therefore, to achieve higher throughput, MongoDB has to use multi-threading to process requests. So we measure the throughput of different workloads in the case of multi-threading (50 threads) and the results are shown in Figure~\ref{mongo2}. In this case, using SmartNIC to expand the host resources does not show obvious performance improvement. It is because a SmartNIC only has eight cores and a large number of threads compete for the limited resources on a single core, which does not scale as well as a host. This result is consistent with our findings in Section \ref{sec3}.

\subsection{Avoid Directly Employing}
\label{secg4}
In \textbf{Guideline 4} we argue that many of the existing design approaches based on on-path SmartNIC cannot be applied to off-path SmartNICs. This is because of the huge difference in characteristics between the two types of SmartNICs. We take Xenic as an example. Xenic uses an on-path SmartNIC as a cache, storing recently accessed data. 
This method is also used in KV-Direct\cite{li2017kv}, which uses FPGA-based SmartNICs. Since the NIC is located between the host core and the network, when the cache hits, the PCIe latency from the NIC to the host core is avoided. In this way, the latency for the client to access data is significantly reduced. Therefore, someone may ask whether this method can be adopted when using an off-path SmartNIC.

Here we implement a key-value cache on the SmartNIC. To demonstrate that our guideline is feasible, we imitate the design in Xenic as closely as possible. When a remote client wants to read data, it first accesses the SmartNIC of the server. If the data to access is on the SmartNIC, the SmartNIC returns the result to the client immediately. If the data to be accessed is not found on the SmartNIC, the SmartNIC reads the data from its host. After obtaining the data, the SmartNIC returns the result to the client.

\begin{figure}
\centering
\subfigure[Average latency.]{
\begin{minipage}[t]{0.47\linewidth}
\centering
\includegraphics[width=\linewidth,trim=20 0 20 0,clip]{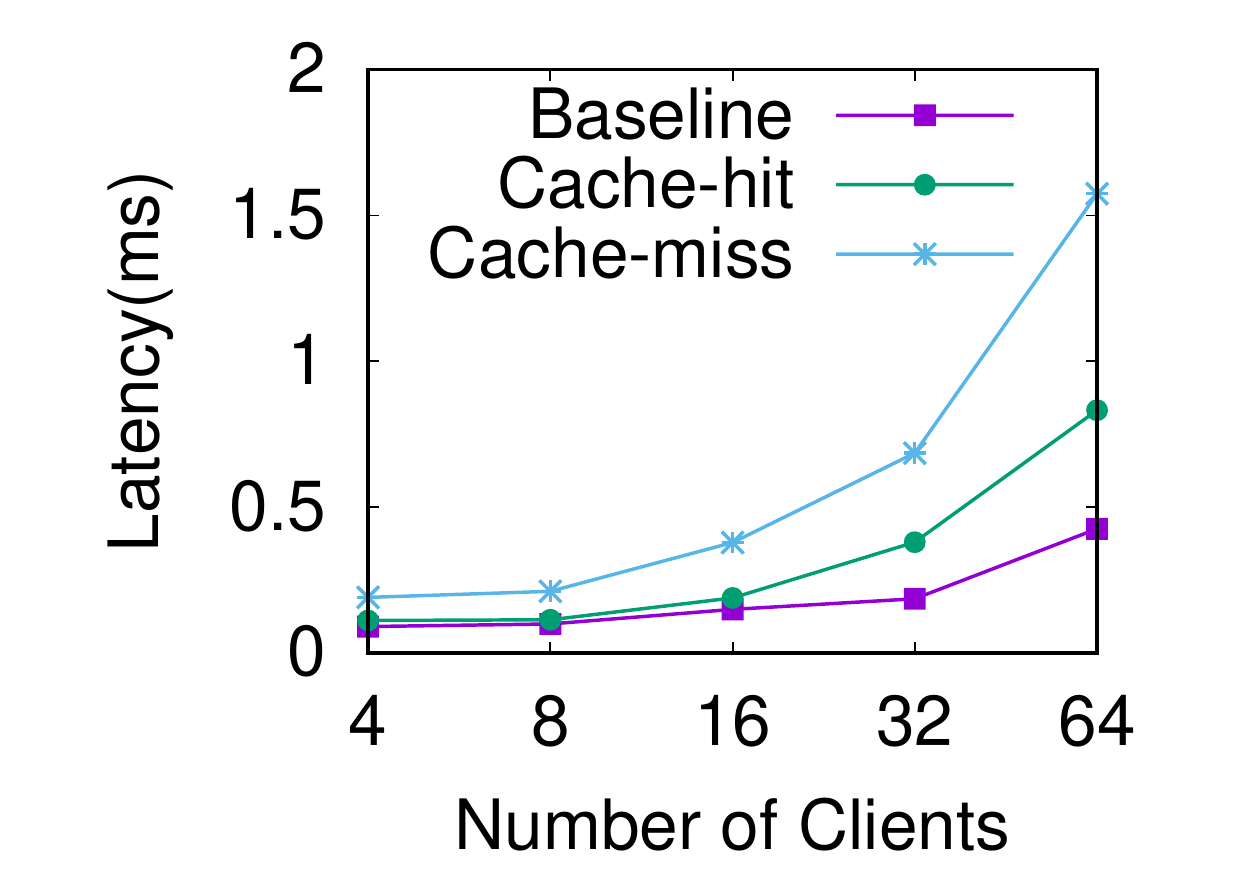}
\end{minipage}%
}%
\subfigure[P99 tail latency.]{
\begin{minipage}[t]{0.47\linewidth}
\centering
\includegraphics[width=\linewidth,trim=20 0 20 0,clip]{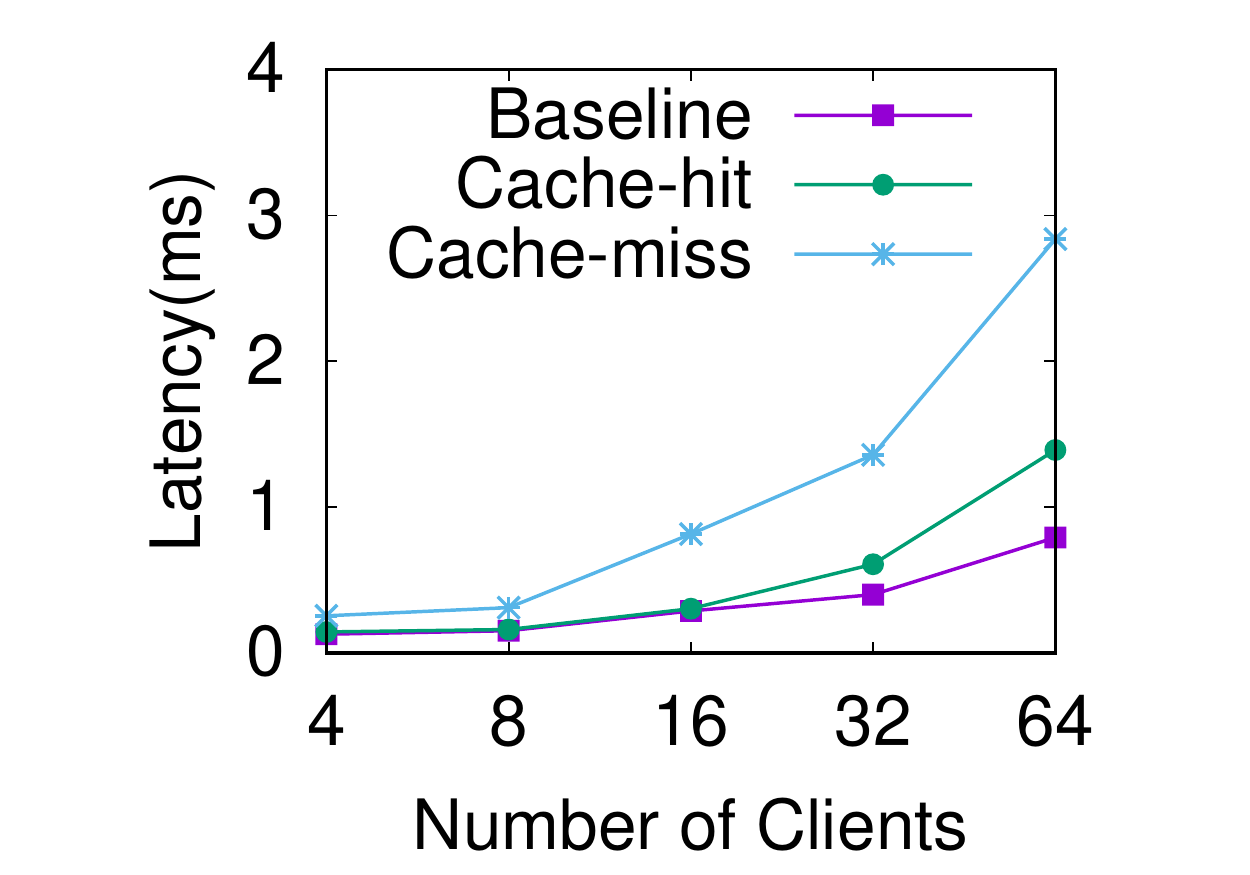}
\end{minipage}
}%

\centering
\caption{The performance when treating the SmartNIC like a Cache.}
\label{figcac}
\end{figure}

We test three situations: all the data is on the host and no SmartNIC is used (Baseline); the SmartNIC is used as a cache and the SmartNIC contains the data to access (Cache-hit); the SmartNIC is used as a cache and the SmartNIC does not have the data to access (Cache-miss). We measure the average and tail latency of GET operations. The results are shown in Figure \ref{figcac}. Using the SmartNIC as a cache does not achieve better performance, no matter the average latency or the tail latency. Among these three cases, the latency of Baseline is the lowest. Cache-hit is slightly higher than Baseline and Cache-miss is much higher than Baseline. This result is different from the performance of systems based on on-path SmartNICs. Even if the cache can be hit all the time, the latency will not be lower than that of not using the SmartNIC. Xenic uses the SmartNIC as a cache to avoid the PCIe latency from the NIC to the host when the cache is hit. On the contrary, applying this idea to an off-path SmartNIC cannot save the latency and it can even make the performance worse. This is because data packets go through the switch on the NIC and the complex network stack when going from the host to the core of the off-path SmartNIC.

The cases of Cache-hit and Cache-miss represent two extreme cases where the cache hit rate is 100\% and 0\%. In general, the latency to access data falls between these two cases. Therefore, regardless of the cache eviction strategy, the latency of the system when using the SmartNIC as a cache will always be higher than the baseline. As a result, such a design would not only increases the complexity of the system but also makes performance worse. This experiment points out that the difference in characteristics of the on-path and off-path SmartNICs have a huge impact on system performance. Simply employing the design method of on-path SmartNICs will not work.

\section{Related Work and Discussion}
\label{secrelated}

\para{On-path SmartNICs.}
At present, most of the studies on multi-core SoC SmartNICs are based on on-path SmartNICs\cite{Floem,liu2019offloading,liu2019e3,qiu2021automated,schuh2021xenic,shashidhara2022flextoe,gao2021ovs}. Although only low-level interfaces can be used to offload applications, on-path SmartNICs provide high offloading performance because they can directly operate on packets. For example, FLOEM\cite{Floem} provides a unified framework which aims at simplifying the development of network applications that are split across host CPUs and NICs. E3\cite{liu2019e3} offloads microservices to SmartNICs and achieves better performance. Xenic\cite{schuh2021xenic} uses asynchronous and aggregated execution models to reduce the network overhead of distributed transaction systems and improve core efficiency. However, according to \textbf{Guideline 4}, their design method is not suitable for off-path SmartNICs. ipipe\cite{liu2019offloading} designs a framework with the goal of maximizing SmartNIC resources utilization. It has a scheduling algorithm that mixes FCFS and DDR which allows tasks to migrate between the host and the SmartNIC constantly. However, this design method is also not applicable to off-path SmartNICs. It is because the huge communication overhead between the SmarNIC and the host will slow down the execution efficiency of the application, even if the processor cores on the SmartNIC are fully occupied.

\para{Off-path SmartNICs.}
With the release of off-path SmartNICs, some researches on off-path SmartNICs appear\cite{tork2020lynx,le2017uno,min2021gimbal,kim2021linefs,fang2021hypernat,sun2022skv}. For example, HyperNAT\cite{fang2021hypernat} offloads network address translation to off-path SmartNICs. SKV\cite{sun2022skv} uses a BlueField SmartNIC to offload data replication in a distributed key-value store. Replication is a latency-insensitive background operation. According to \textbf{Guideline 2} it is suitable to offload. SKV exceeds the performance of the baseline in the write load test, but there is no performance improvement in the face of the read load. It is because the replication operation is only triggered when writing. Only in this situation the offloading can reduce CPU consumption. LineFS\cite{kim2021linefs} is a SmartNIC-offloaded distributed file system. LineFS offloads replication and publishing to the BlueField SmartNIC and uses pipelines to accelerate. Pipelining can guarantee higher throughput, but cannot obviously reduce the latency. Fortunately, these two operations are latency-insensitive background operations, thus complying with \textbf{Guideline 2}. However, the design of LineFS does not significantly improve the throughput when the host is relatively idle. Only when the streamcluster stress testing tool is executed on the host at the same time, the offloading mechanism of LineFS can reduce CPU competition and bring better throughput. In addition, in the evaluation of latency, the average latency and P99 tail latency of LineFS are higher than the baseline. Only the P999 tail latency of LineFS is better than the baseline when executing streamcluster. In such a design, only a limited performance improvement is obtained, which shows that offloading with an off-path SmartNIC is still very challenging.

There are also works based on off-path SmartNICs, but they do not take advantage of SmartNIC offloading. Lynx\cite{tork2020lynx} implements a framework for scheduling and managing heterogeneous AI accelerators on the BlueField SmartNIC, which exceeds the performance of the traditional host-center method. However, this framework can also run on a host. Their experiments show that the throughput of Lynx running on the SmartNIC is lower than that of Lynx running on the host. This shows that the framework works but running it on an SmartNIC is probably not the best option. Gimbal\cite{min2021gimbal} designs a software storage switch on SmartNIC that enables multi-tenant storage disaggregation. In their experiments, Gimbal and all other baselines run on SmartNICs. Gimbal outperforms other baselines, but they have not evaluated the performance of Gimbal on the host. So it is unclear whether it will perform better on an SmartNIC than on a host. These works dealt with SmartNICs, but neither focuses on how to take full advantage of SmartNIC offloading. They just run some components on the SmartNIC because the complete operating system on the SmartNIC can execute programs just like general hosts.

\section{Conclusion}\label{sec:conclusion}
This paper provides a systematic study on how to best utilize off-path multi-core SoC SmartNICs. We take BlueField as an example to analyze the internal structure of an off-path SmartNIC, and evaluate its performance with some microbenchmarks. After that, we present several guidelines for using off-path SmartNICs, which provide some methods on how to leverage the accelerators and general-propose cores on the SmartNICs. We apply these guidelines to several cases, including a regular expression matching application, a key-value store, and a document storage system. By using these guidelines, the performance of these systems is improved significantly. We hope these design guidelines can help developers to integrate SmartNICs into their systems.

\bibliographystyle{unsrt}
\bibliography{sample}
\end{document}